\documentclass[]{article}
\usepackage{amsmath, amsthm, amssymb,amsbsy, appendix, bm, graphicx, mathrsfs,enumerate,indentfirst,array,tabularx,booktabs,float,geometry,exscale,relsize,graphicx,subfig,fontenc,inputenc,authblk,natbib,threeparttable,setspace}
\usepackage[ruled,linesnumbered]{algorithm2e}
\usepackage[figuresright]{rotating}
\newtheorem{theorem}{Theorem}

\setcitestyle{authoryear,round}
\usepackage{hyperref}
\hypersetup{
    colorlinks=true,
    allcolors = {blue}
}

\title{\textbf{Semiparametric marginal promotion time cure model for clustered survival data}}
\author[a]{Fei Xiao}
\author[b,c]{Yingwei Peng}	
\author[d]{Dipankar Bandyopadhyay}	
\author[a]{Yi Niu \thanks{Corresponding author: niuyi@dlut.edu.cn}}
\affil[a]{School of Mathematical Sciences, Dalian University of Technology, Dalian, Liaoning, China}
\affil[b]{Department of Public Health Sciences,
    Queen's University, Kingston, Ontario, Canada}
\affil[c]{Department of Mathematics and Statistics,
    Queen's University, Kingston, Ontario, Canada}
\affil[d]{Department of Biostatistics, Virginia Commonwealth University, Richmond, Virginia, USA}

\begin{document}
    \date{}
    \maketitle
    \thispagestyle{empty}

    \begin{abstract}
        \setlength{\parindent}{2em}\setlength{\parindent}{2em}
        Modeling clustered/correlated failure time data has been becoming increasingly important in clinical trials and epidemiology studies.
        In this paper, we consider a semiparametric marginal promotion time cure model for clustered right-censored survival data with a cure fraction.
        We propose two estimation methods based on the generalized estimating equations and the quadratic inference functions
        and prove that the regression estimates from the two proposed methods are consistent and asymptotic normal and that the estimates from the quadratic inference functions are optimal.
        The simulation study shows that the estimates from both methods are more efficient than those from the existing method no matter whether
        the correlation structure is correctly specified.
        The estimates based on the quadratic inference functions achieve higher efficiency compared with those based on the generalized estimating equations under the same working correlation
        structure.
        An application of the proposed methods is demonstrated with periodontal disease data and new findings are revealed in the analysis.

    \end{abstract}

    \section{Introduction} \label{section:1}
    Clustered and multivariate failure time data are commonly encountered in biomedical studies.
    For example, in a study of infections in kidney patients \citep{MA1991BTC}, several infections may be observed
    in each patient. If the survival time is defined as the time between the end of the previous infection and the beginning of the current infection, the times to multiple recurrences of infections on each patient form a cluster and may be correlated to one another. Although it is reasonable to assume that the recurrent event times from the different subjects are independent, the correlation between the event times from the same subject should not be ignored.
    To appropriately account for the correlation among the clustered survival times, the two most studied approaches are marginal models and frailty models.
    The marginal models focus on the population average on the marginals of the joint distribution of data from one cluster, and the correlation is often treated as a nuisance parameter in the model to reduce the dependence of the marginal models on the specification of the unobservable correlation structure of clustered data. Alternatively, the frailty models explicitly formulate the underlying dependence structure by random effects, and the failure times are assumed to be independent conditional on the unobservable frailty.

    Another common issue in studies with failure time data is that a fraction of study subjects with censored survival times may never experience the event of interest however long the follow-up is and they may be considered cured. For instance, in clinical studies of some early-stage cancers, the Kaplan-Meier  survival curves for disease-free survival times often show a long and stable plateau with heavy censoring at the tail, indicating that the long-term censored subjects with longer survival times than uncensored times are likely to be cured.
	The commonly used survival models to accommodate the possible cure fraction in survival data in the literature are the two-component mixture cure (MC) model \citep{Berkson1952JASA,Boag1949JRSSB} and the promotion time cure (PTC) model \citep{Yakovlev1996BOOK,Yin2005CJOS}. The latter is also known as the non-mixture cure model \citep{Li2001SPL}, the bounded cumulative hazard model \citep{Tsodikov2003JASA}, or the proportional hazards cure model \citep{Ma2008JASA}.

	Both the marginal method and the frailty method have been extensively investigated with the MC model for analyzing clustered survival data with a cure fraction. \cite{Peng2007LDA} and \cite{Yu2008CSDA} estimated the parameters in marginal MC models using methods proposed by \citet{Peng2000BTC} and \citet{Sy2000BTC} by ignoring the correlation and adjusting the variances of the estimates due to the correlation using the sandwich or Jackknife methods.
    \citet{Niu2013SIM,Niu2014JMVA} and \citet{Niu2018BJ} proposed a generalized estimating equations approach to estimate the parameters in the marginal MC models with a working correlation structure.
    \citet{Su2019SIM} investigated marginal MC models by utilizing the Archimedean copula to assess the strength of association within clusters, which extended the work of \citet{Chatterjee2001BTC} and \citet{Wienke2003BTC} with covariate effects.
    \citet{Chen2012CSDA} extended the work of \citet{Peng2007LDA} by considering a marginal transformation model for uncured patients and proved the asymptotic properties of the proposed estimators.
    \citet{Shen2021CIS} considered a marginal semiparametric transformation cure model for gap times between recurrent events.
	For the frailty/random effects method, \cite{Yau2001SIM} and \cite{Lai2008SIM} considered two normally distributed random effects in an MC model to accommodate
    the dependence among the cure statuses and among the survival times of uncured subjects and estimated parameters in the model using the best linear unbiased predictor method and the residual maximum likelihood estimator when the baseline survival function is either parametrically or nonparametrically
    specified.
    \citet{Peng2011SIM} suggested a numerical integration method to obtain the maximum likelihood estimates of the parameters in the model while \citet{Rondeau2013SMIMR} compared several forms of random effects MC models with a piecewise constant baseline hazard function.
    \citet{Xiang2011SIM} extended the random effects MC model to clustered interval-censored survival data.
	Other types of random effects were also considered.
    \citet{Schnell2015JRSSC} considered linear combinations of positive stable
    random effects;
    \citet{Tawiah2020SMIMR} incorporated multivariate and time-varying frailties into the MC model;
    \citet{Liu2023CIS} considered a generalized Birnbaum-Saunders frailty distribution \citep{Birnbaum1969JAP}.
	\citet{Cooner2006SMIMR} and \citet{Sepp2014JRSSC} considered hierarchical mixture cure models with random effects in a Bayesian framework.

    Compared to the abundance of work on MC models, the work on PTC models for analyzing clustered failure time data with a cure fraction is sparse in the literature.
    \citet{Herring2002BioS} proposed normal random effects PTC models with nonignorable missing covariates.
    \citet{Banerjee2004BTC} incorporated spatial frailties in parametric PTC models to allow for spatial correlation in the interval-censored survival data.
    A PTC model using a positive stable frailty term to measure the correlation within a cluster was considered by \citet{Chen2002JRSSC}.
    \citet{Yin2008SIM} further extended this work by considering a transformation cure model with a frailty for multivariate failure time data.
    \citet{Lopes2012CSDA} presented both Bayesian and classical estimation approaches for a parametric PTC model with random effects.
    \citet{LD2016BJ} extended the non-mixture cure model by imposing a Gaussian processes prior on the nonlinear structure of continuous covariate for modeling spatially correlated survival data.
    \citet{Yang2021SIM} proposed a class of semiparametric frailty transformation non-mixture cure models for modeling clustered interval-censored failure time data with a cure subgroup.
    All the works above focus on the frailty method within the Bayesian paradigm for modeling correlated survival data. The only work based on the marginal approach we are aware of is the work of \citet{Lam2021SIM} on a marginal transformation PTC model for clustered current status data with informative cluster size. Given the flexibility in specifying the correlation structures in correlated survival time in estimation methods for a marginal model and the easier interpretation of the marginal effects than the conditional effects in a random effects model, it will be useful to investigate estimation methods for marginal models based on the PTC model and provide a viable alternative way to the random effects approach for clustered survival data with a cured fraction when the marginal PTC model is preferred.

    In this paper, we propose a marginal PTC model for analyzing clustered survival data with long-term survivors. Generalized estimating equations (GEE) and quadratic inference functions (QIF) based on the generalized method of moments are developed to estimate the regression parameters and the dependence within clusters is prespecified by working correlation structures in the model, and the proposed estimation method is shown to be optimal among a class of estimating equations.
    We discuss the asymptotic properties of the estimates from the proposed methods and provide valid variance estimation formulas.
    The paper is organized as follows. In Section \ref{section:2}, we introduce the marginal PTC model for clustered survival data with a cured fraction and propose the estimation methods based on the GEE and the QIF to estimate the marginal PTC model. The asymptotic properties of the estimates are discussed in this section too. In Section \ref{section:3}, we conduct simulation studies to evaluate the performance of the proposed estimation methods, including the efficiency gain and the robustness of the methods when the working correlation structure is misspecified. An illustration of the proposed methods with periodontal disease data is given in Section \ref{section:4}. Conclusions and discussions are presented in Section \ref{section:5}.

    \section{Marginal PTC model and estimation methods} \label{section:2}
    \subsection{Marginal PTC model}  \label{section:2.1}
    Suppose that there are $K$ clusters in the clustered failure time data and $n_i$ observations in the $i$th cluster for $i=1,\ldots,K$.
    The total number of observations $N=\sum_{i=1}^{K}n_i$. Let $\tilde{T}_{ij}$ and $C_{ij}$ be the $j$th failure and censoring times in
	the $i$th cluster, and $Y_{ij}$ be the cure status of the $j$th time in the $i$th cluster with $Y_{ij} = 0$ if the subject is cured and 1 otherwise. We assume that $\tilde{T}_{ij}<\infty$ if $Y_{ij} = 1$ and $\tilde{T}_{ij}\equiv\infty$ if $Y_{ij} = 0$ (the cure threshold $\infty$ can be replaced with a finite value denoted as $\tau$ in practice. See more discussions in Section~\ref{procedure}). The observed failure time is $T_{ij}=\min(\tilde{T}_{ij},C_{ij})$ and the censoring indicator is $\delta_{ij}=I(\tilde{T}_{ij}\leq C_{ij})$, where $I(A)$ is 1 if $A$ is true and 0 otherwise. Let $\bm{X}^{*}_{ij}$ be a vector of $p_X$ covariates that may have effects on the failure time distribution and $\bm{X}_{ij}=(1,\bm{X}^{*T}_{ij})^T$ is full rank.
    Given $\bm{X}^{*}_{ij}$, we assume that $\tilde{T}_{ij}$ and $\tilde{T}_{ij'}$ are correlated for $j\neq j'$, $\tilde{T}_{ij}$ and $\tilde{T}_{i'j'}$ are independent for $i\neq i'$, $C_{ij}$ is independent of $\tilde{T}_{ij}$, and the marginal survival function of $\tilde{T}_{ij}$ follows the PTC model:
    \begin{equation}
        S(t;\bm{X}_{ij})=P(\tilde{T}_{ij}>t;\bm{X}_{ij})=\exp(-\alpha (\bm{X}_{ij})F_{\mathfrak{p}}(t)),   \label{1}
    \end{equation}
    where $F_{\mathfrak{p}}(t)$ is an unknown proper baseline cumulative distribution function and $\alpha(\cdot)$ is a prespecified positive link function. Here we consider $\alpha(\bm{X}_{ij}) = \exp(\bm{\beta}^T \bm{X}_{ij})$ with unknown parameters $\bm{\beta}= (\beta_0,\beta_{1},\ldots,\beta_{p_{X}})^{T}$.
    Given $\bm{X}^{*}_{ij}$, the cure probability under this model is given by
	\begin{gather}
		P(Y_{ij} = 0) = P(\tilde{T}_{ij}=\infty)=\lim_{t\to\infty}S(t ;\bm{X}_{ij})=\lim_{t\to\infty} \exp(-e^{\bm{\beta}^T\bm{X}_{ij}}F_{\mathfrak{p}}(t)) =\exp(-e^{\bm{\beta}^T\bm{X}_{ij}}).
		\label{cure}
	\end{gather}

    \subsection{The GEE method for estimating \protect\boldmath{${\beta}$}}  \label{section:2.2}
    Let $\bm\theta=(\bm{\beta},F_{\mathfrak{p}}(t))$ denote the unknown parameters to be estimated in \eqref{1}.
    If we ignore the dependence within clusters, given $F_{\mathfrak{p}}(t)$,
    the unknown parameters $\bm{\beta}$ in the model \eqref{1} can be estimated based on the following log-likelihood function with the observations $O= \{(t_{ij},\delta_{ij},\bm{X}_{ij}),i = 1,2,\cdots,K, j = 1,2,\cdots, n_i \}$, i.e.,
    \begin{equation}
        \begin{aligned}[b]
            l(\theta;O)&=\log\prod\limits_{i=1}^K\prod\limits_{j=1}^{n_i}h(t_{ij};\bm{X}_{ij})^{\delta_{ij}}S(t_{ij};\bm{X}_{ij}) \\
            &=\log\prod\limits_{i=1}^K\prod\limits_{j=1}^{n_i}\left\{f_\mathfrak{p}(t_{ij})\exp({\bm{\beta}^T}\bm{X}_{ij})\right\}^{\delta_{ij}}\exp\left\{-F_\mathfrak{p}(t_{ij})\exp(\bm{\beta}^T\bm{X}_{ij})\right\} \\
            &=\log\prod\limits_{i=1}^K\prod\limits_{j=1}^{n_i}\left\{[\exp({\bm{\beta}^T}\bm{X}_{ij})]^{\delta_{ij}}\exp[-F_\mathfrak{p}(t_{ij})\exp({\bm{\beta}^T}\bm{X}_{ij})]\right\}+\log\prod\limits_{i=1}^K\prod\limits_{j=1}^{n_i}f^{\delta_{ij}}_\mathfrak{p}(t_{ij}) \label{2}
        \end{aligned}
    \end{equation}
    where $h(t;\bm{X}_{ij})$ is the hazard function corresponding to $S(t;\bm{X}_{ij})$, $f_{\mathfrak{p}}(t)$ is the probability density function corresponding to $F_{\mathfrak{p}}(t)$,
    and the score function of $\bm{\beta}$ based on \eqref{2} is,
     \begin{equation}
		 U^I({\bm{\beta}})=\sum\limits_{i=1}^K U^I_i(\bm{\beta})=\sum\limits_{i=1}^K\left\{\frac{\partial{\bm{\mu}}(\bm{X}_i)}{\partial{\bm{\beta}}}\right\}^T\bm{B}_i^{-1}\left\{\bm{\delta_i}-\bm{F}_\mathfrak{p}(\bm{t}_{i})\bm{\mu}(\bm{X}_i)\right\},\label{3}
    \end{equation}
	where ${\bm{\mu}}(\bm{X}_i)=(\mu(\bm{X}_{i1}),\ldots,\mu(\bm{X}_{in_i}))^T=(\exp(\bm{\beta}^T\bm{X}_{i1}),\ldots,\exp(\bm{\beta}^T\bm{X}_{in_i}))^T$, $\bm{B}_i=\operatorname{diag}\{\mu(\bm{X}_{i1}),\ldots,\mu(\bm{X}_{in_i})\}$, $\bm{\delta}_i=(\delta_{i1}, \delta_{i2},\ldots,\delta_{in_i})^T$,
	and $\bm{F}_\mathfrak{p}(\bm{t}_{i})=(F_\mathfrak{p}(t_{i1}),F_\mathfrak{p}(t_{i2}),\ldots,F_\mathfrak{p}(t_{in_i}))^T$. The estimate of $\bm{\beta}$ based on $U^I(\bm{\beta})=0$ is denoted as $\hat{\bm{\beta}}_{I}$.

    Because of the potential correlation within clusters, the estimator $\hat{\bm{\beta}}_{I}$ based on \eqref{3} may not be efficient, even though the marginal model is correctly
    specified \citep{Peng2007LDA,Yu2008CSDA}. To improve the estimation efficiency, following the idea of the seminal work of \citet{Liang1986BTK},
    we propose a weighted generalized estimating function of $\bm{\beta}$,
    i.e.,
    \begin{equation}
		U^G(\bm{\beta})=\sum\limits_{i=1}^K U^G_i(\bm{\beta})=\sum\limits_{i=1}^K\left\{\frac{\partial{\bm{\mu}}(\bm{X}_i)}{\partial\bm{\beta}}\right\}^T\left\{\bm{B}_i^{1/2}\bm{Q}_i({\bm{\rho}})\bm{B}_i^{1/2}\phi\right\}^{-1}\bm{W}_i\left\{\bm{\kappa}_i-\bm{\mu}(\bm{X}_i)\right\} ,\label{4}
    \end{equation}
	 where $\bm{W}_i=\operatorname{diag}\{F_\mathfrak{p}(t_{i1}),\dots,F_\mathfrak{p}(t_{in_i})\}$, $\bm{\kappa}_i=(\kappa_{i1},\dots,\kappa_{in_i})^T=(\delta_{i1}/F_\mathfrak{p}(t_{i1}),\dots,\delta_{in_i}/F_\mathfrak{p}(t_{in_i}))^T$, and $\bm{Q}_i({\bm{\rho}})=(q_{jj'}(\bm\rho))_{n_i\times n_i}$ is a working correlation matrix specified up to a few unknown parameters in $\bm{\rho}$ and is used to approximate the correlation among the observations within the $i$th cluster.
     Some commonly used working correlation structures include the exchangeable (equicorrelated, compound symmetric) structure with $q_{jj'}(\bm\rho)=\rho$ for $j\neq j'$ and 1 otherwise, the first-order autoregressive (AR(1)) structure with $q_{jj'}(\bm\rho)=\rho^{|j-j'|}$ for $j\neq j'$ and 1 otherwise, and the unstructured correlation structure with $q_{jj'}(\bm\rho)=\rho_{jj'}$ for $j\neq j'$ and 1 otherwise.
     The scale parameter $\phi$ is used to accommodate the over- or under-dispersion, and estimated by $\hat{\phi}=\sum_{i=1}^{K}\sum_{j=1}^{n_{i}}\hat{e}_{i j}^{2}/(N-p_{X}-1)$ with the standardized Pearson residuals
    $\hat{e}_{ij} =\{\kappa_{ij}-\mu\left(\bm{X}_{ij}\right)\}/\{\mu\left(\bm{X}_{ij}\right)\}^{\frac{1}{2}}$. The form of estimator $\hat{\bm\rho}$ depends on the structure of working correlation matrix $Q_i(\cdot)$. For example,  $\hat{\rho}=\phi^{-1} \sum_{i=1}^{K}\sum_{j\neq k}\hat{e}_{ij}\hat{e}_{ik} / (\sum_{i=1}^{K}n_{i}(n_{i}-1)-p_X-1)$ for the exchangeable structure, and $\hat{\rho}=\phi^{-1} \sum_{i=1}^{K}\sum_{j\leq n_{i}-1}\hat{e}_{ij}\hat{e}_{i,j+1} / (\sum_{i=1}^{K}(n_{i}-1)-p_X-1)$ for the AR(1) structure.
    It is worth pointing out that $U^G(\bm{\beta})$ reduces to $U^I(\bm{\beta})$ when $Q_i({\bm{\rho}})$ is specified as an identity matrix and $\phi=1$.
    The estimate of $\bm\beta$ based on $U^G(\bm{\beta})=0$ is denoted as $\hat{\bm\beta}_{G}$.

    Let $\bm{\beta}^*$ be the true value of $\bm\beta$. We can establish the consistency and asymptotic normality of $\hat{\bm{\beta}}_G$ as follows.
    \begin{theorem}  \label{thm:Theorem-1}
        Under some regularity conditions,
		\begin{enumerate}[(a)]
        \item $\hat{\bm{\beta}}_{G}$ is a consistent estimate of $\bm{\beta}^*$,
        
       \item $\sqrt{K} (\hat{\bm{\beta}}_{G}-\bm{\beta}^*) \to N(0,\Sigma_{G}(\bm{\beta}^*)) $ (in distribution) as $K \to \infty$, where
        $$ \Sigma_{G}(\bm{\beta}^*) = \left( -\frac{\partial U(\bm{\beta}^*)}{\partial \bm{\beta} }\right) ^{-1}\left( \sum _{i=1}^{K} U_{i}(\bm{\beta}^*)U_{i}(\bm{\beta}^* )^{T}\right) \left( -\frac{\partial U(\bm{\beta}^* )}{\partial \bm{\beta} }\right) ^{-T},  $$
        and  $\Sigma_{G}(\bm{\beta}^*)$ can be consistently estimated by $\hat{\Sigma}_{G}=\Sigma_{G}(\hat{\bm{\beta}}_{G}).$
      \end{enumerate}
      \end{theorem}
    Proof of the Theorem \ref{thm:Theorem-1} is presented in Appendix A.

    \subsection{The QIF method for estimating \protect\boldmath{$\beta$}} \label{section:2.3}
	 As pointed out by \citet{Crowder1995BTK}, when the specification of working correlation structure $Q_i(\bm\rho)$ in \eqref{4} is incorrect, the moment estimator of the nuisance parameter ${\bm\rho}$ no longer leads to the optimal estimation of $\bm{\beta}$. To address this issue, we consider using a quadratic inference function (QIF) \citep{Qu2000BTK} to improve the efficiency of $\hat{\bm\beta}_G$. The feature of the QIF method is that it does not need to estimate the unknown correlation parameters and thus relaxes the dependence of the estimation of $\bm\beta$ on $\hat{\bm\rho}$. Suppose that the inverse of the working correlation matrix $\bm{Q}_i(\bm{\rho})$ can be expressed as a linear combination of $m(m>1)$ basis matrices. That is
    \begin{equation}
		\phi^{-1}\bm{Q}_i^{-1}(\bm{\rho})=z_1\bm{M}_1+\cdots+z_m\bm{M}_m, \label{8}
    \end{equation}
	where $\bm{M}_1,\ldots, \bm{M}_m$ are known basis matrices with elements only 0 and 1, and $z_1,..., z_m$ are unknown parameters.
	Then the estimating function $U^G(\bm\beta)$ in \eqref{4} can be rewritten as
        \begin{align}
			U^G(\bm{\beta})& =\sum_{i=1}^{K}\left\{\frac{\partial{\bm\mu}(\bm{X}_{i})}{\partial{\bm\beta}}\right\}^{T}\bm{B}_{i}^{-1/2}\{z_{1}\bm{M}_{1}+\cdots+z_{m}\bm{M}_{m}\}\bm{B}_{i}^{-1/2}\bm{W}_{i}\{\bm{\kappa}_{i}-\bm{\mu}(\bm{X}_{i})\}  \nonumber\\
			&=\sum_{s=1}^{m}\bm{Z}_{s}\sum_{i=1}^{K}\left\{\frac{\partial{\bm\mu}({\bm{X}}_{i})}{\partial{\bm\beta}}\right\}^{T}\bm{B}_{i}^{-1/2}\bm{M}_{s}\bm{B}_{i}^{-1/2}\bm{W}_{i}\left\{{\bm\kappa}_{i}-{\bm\mu}({\bm{X}}_{i})\right\}, \label{9}
        \end{align}
    where
	$$\bm{Z}_1=\begin{bmatrix}z_1&0&\cdots&0\\ 0&\ddots&\ddots&\vdots\\ \vdots&\ddots&\ddots&0\\ 0&\cdots&0&z_1\end{bmatrix}_{(p_X+1)\times (p_X+1)}\quad,\cdots,\bm{Z}_m=\begin{bmatrix}z_m&0&\cdots&0\\ 0&\ddots&\ddots&\vdots\\ \vdots&\ddots&\ddots&\vdots\\ 0&\cdots&0&z_m\end{bmatrix}_{(p_X+1)\times (p_X+1)}.$$\par
   We further define the extended score vector $G_K(\bm\beta)$ as
        \begin{align}
            G_K(\bm{\beta})=& \frac{1}{K}\sum\limits_{i=1}^K g_i(\bm{\beta}) \nonumber \\
			=&\frac{1}{K}\sum\limits_{i=1}^K \begin{bmatrix}\left\{\frac{\partial {\bm\mu}(\bm{X}_i)}{\partial{\bm\beta}}\right\}^T \bm{B}_i^{-1/2}\bm{M}_1\bm{B}_i^{-1/2}W_i\{\bm{\kappa}_i-\bm{\mu}(\bm{X}_i)\}\\ \vdots\\ \left\{\frac{\partial{\bm\mu}(\bm{X}_i)}{\partial{\bm\beta}}\right\}^T \bm{B}_i^{-1/2}\bm{M}_m\bm{B}_i^{-1/2}W_i\{\bm{\kappa}_i-\bm{\mu}(\bm{X}_i)\}\end{bmatrix}_{m(p_X+1)\times 1}. \label{10}
        \end{align}
	It is easy to show that $U^G(\bm{\beta})=K(\bm{Z}_1\cdots \bm{Z}_m)G_K(\bm{\beta})$ and the solution of $G_K(\bm{\beta}) = 0$ is also the solution of $U^G(\bm{\beta}) = 0$ without considering the unknown parameter ${z} = (z_1,..., z_m)^T $. However, the dimension of $G_K(\bm{\beta})$ is greater than that of the unknown parameters $\bm{\beta}$. Thus, we consider estimating $\bm{\beta}$ by minimizing a QIF of $\bm{\beta}$, which is a combination of the estimating equations in $G_K(\bm{\beta})$ using the generalized method of moment (GMM) \citep{Hansen1982Eco} and is given by $G^T_K(\bm{\beta})C_K^{-1}(\bm{\beta})G_K(\bm{\beta})$, where $C_K(\bm{\beta})=(1/K)\sum_{i=1}^{K}g_i(\bm{\beta})g_i^T(\bm{\beta})$ is an estimate of the covariance matrix of $G_K(\bm{\beta})$.
	This method tends to give less weight to the element of $G_K(\bm\beta)$ that has a larger variance.
	The corresponding estimating function of $\bm{\beta}$ is
    $$\begin{aligned}
        \frac{\partial}{\partial{\bm\beta}}G^T_K(\bm{\beta})C_K^{-1}(\bm{\beta})G_K(\bm{\beta})&=2\dot{G}_K^T(\bm{\beta})C_K^{-1}(\bm{\beta})G_K(\bm{\beta})-G_K^T(\bm{\beta})C_K^{-1}(\bm{\beta})\dot{C}_K(\bm{\beta})C_K^{-1}(\bm{\beta})G_K(\bm{\beta})\\
        &=2\dot{G}_K^T(\bm{\beta})C_K^{-1}(\bm{\beta})G_K(\bm{\beta})-O_p(K^{-1}),
    \end{aligned}$$
    where $\dot{G}_K(\bm\beta)=\partial{G_K(\bm\beta)}/\partial{\bm\beta}$ and $\dot{C}_K(\bm\beta)=\{\partial{C_K(\bm\beta)}/\partial{\beta_0},\cdots,\partial{C_K(\bm\beta)}/\partial{{\beta}_{p_X}}\}$.
    Since $\frac{\partial}{\partial\bm\beta}G^T_K(\bm{\beta})C_K^{-1}(\bm{\beta})G_K(\bm{\beta})$ can be asymptotically approximated by $2\dot{G}_K^T(\bm{\beta})C_K^{-1}(\bm{\beta})G_K(\bm{\beta})$, a new estimating function for $\bm{\beta}$ based on the QIF can be expressed as
    \begin{equation}
	U^Q(\bm{\beta})=\frac{1}{K}\sum_{i=1}^{K}U^{Q}_{i}(\bm{\beta})=\frac{1}{K}\sum_{i=1}^{K}\dot{G}_K^T(\bm{\beta})C_K^{-1}(\bm{\beta})g_i(\bm{\beta}).  \label{UQ}
    \end{equation}
    The estimate of $\bm\beta$ based on $U^Q(\bm{\beta})=0$ is denoted as $\hat{\bm\beta}_{Q}$.

    The proposed estimator $\hat{\bm\beta}_{Q}$ enjoys some desirable properties, such as consistency, asymptotic normality, smaller variance than $\hat{\bm\beta}_{G}$, and optimal efficiency. Details are given in the following theorem:
    \begin{theorem}  \label{thm:Theorem-2}
        Under some regularity conditions,
		\begin{enumerate}[(a)]
        \item $\hat{\bm{\beta}}_{Q}$ is a consistent estimator of $\bm{\beta}^*$,
        \item $\sqrt{K} (\hat{\bm{\beta}}_{Q}-\bm{\beta}^*) \to N(0,\Sigma_{Q}(\bm{\beta}^*) ) $ (in distribution) as $K \to \infty$, where
        $$\Sigma_{Q}(\bm{\beta}^*)=\left[\dot{G}_K^T(\bm{\beta}^*)C_K^{-1}(\bm{\beta}^*)\dot{G}_K(\bm{\beta}^*)\right]^{-1}\Omega_{Q}(\bm{\beta}^*)\left[\dot{G}_K^T(\bm{\beta}^*)C_K^{-1}(\bm{\beta}^*)\dot{G}_K(\bm{\beta}^*)\right]^{-T},  $$
        $$\Omega_{Q}(\bm{\beta}^*) =\sum U^{Q}_{i}(\bm{\beta}^*)U^{Q}_{i}{(\bm{\beta}^*)}^{T} ,$$
        and $\Sigma_{Q}(\bm{\beta}^*)$ can be consistently estimated by $\hat{\Sigma}_{Q}=\Sigma_{Q}(\hat{\bm{\beta}}_{Q}).$
	\item $\hat{\bm\beta}_Q$ has a smaller asymptotic variance than $\hat{\bm\beta}_G$. That is, $\Sigma_Q \le \Sigma_G$.
	\item $\hat{\bm\beta}_Q$ is an optimal estimator in the sense that the asymptotic covariance of $\hat{\bm\beta}_Q$ reaches the minimum among the estimators from a class of estimating equations $U^E(\bm{\beta})=0$, where
    \begin{equation}
		U^{E}(\bm{\beta})=\sum _{s=1}^{m}\bm{E}_{s}\sum _{i=1}^{K}\left\{ \frac{\partial \bm{\mu} (\bm{X}_{i})}{\partial \bm{\beta} }\right\} ^{T}\bm{B}_{i}^{-1/2}\bm{M}_{s}\bm{B}_{i}^{-1/2}\bm{W}_{i}\left\{ \bm{\kappa} _{i}-\bm{\mu} (\bm{X}_{i})\right\},   \label{14-10}
    \end{equation}
	where $\bm{E}_{s}$ $(s=1,\ldots ,m)$ are the $(p_X+1)\times (p_X+1)$ arbitrary nonrandom matrices.
        \end{enumerate}
    \end{theorem}
    Proof of the Theorem \ref{thm:Theorem-2} is given in Appendix B.

	\subsection{The complete estimation procedure}\label{procedure}

	The above discussions about estimating $\bm\beta$ are based on a known baseline cumulative distribution function $F_{\mathfrak{p}}(t)$. Like in the Cox model, $F_{\mathfrak{p}}(t)$ is a nuisance parameter in the model, and the cure probability in \eqref{cure} does not depend on $F_{\mathfrak{p}}(t)$. However, $F_{\mathfrak{p}}(t)$ needs to be estimated from data when estimating $\bm\beta$.
	To estimate $F_{\mathfrak{p}}(t)$ consistently, we consider the nonparametric maximum likelihood estimator $\hat{F}_{\mathfrak{p}}(t)$ using the Lagrange multiplier method proposed by \citet{Portier2017BNL} and \citet{Zeng2006JASA} under the independent observation assumption, which is given by
     \begin{equation}   \label{FP}
      \hat{F}_{\mathfrak{p}}(t)=\frac{1}{N}\sum_{i=1}^{K}\sum _{j=1}^{n_i} \frac{\delta_{ij}I{(T_{ij}\leq t)}}{ \hat{R}_{\bm{\beta}}(T_{ij})-\hat{\lambda}_{\bm{\beta}}},
  \end{equation}
	where $\hat{\lambda}_{\bm{\beta}}$ satisfies $\hat{F}_{\mathfrak{p}}(+\infty)=1$, and $\hat{R}_{\bm\beta}(u) = \frac{1}{N}\sum_{l=1}^{K}\sum _{s=1}^{n_i}\exp \left( \bm{\beta}^{T}\bm{X}_{ls} \right) (I(u\leq T_{ls}\leq \tau)+I(T_{ls}>\tau))$ with a known cure threshold $\tau$ which should be larger than the largest uncensored observation and justifiable with the subject knowledge in the related field.
	This estimator is a step-function with a finite number $\sum_{i=1}^{K}\sum _{j=1}^{n_i}\delta_{ij}$ of bounded jumps that sum up to one.
	Other estimation methods for $F_{\mathfrak{p}}(t)$ include the restricted nonparametric maximum likelihood estimation method \citep{Tsodikov2002SIM,Tsodikov2003JASA} and a Bernstein-based sieve empirical likelihood method \citep{Lam2021SIM,Han2021BIO}.

   With the estimator of $F_{\mathfrak{p}}(t)$, we can summarize the proposed estimation procedure for $\bm\beta$ based on the GEE or QIF methods in the following steps.
	
	\begin{enumerate}[{Step }1.]
	\item Set the initial value for $\bm{\beta}$ and calculate $\hat{F}_{\mathfrak{p}}(\cdot)$ based on \eqref{FP}.
	\item Given $\hat{F}_{\mathfrak{p}}(\cdot)$, update $\hat{\bm{\beta}}_Q$ with \eqref{UQ} using the Newton-Raphson method or
	\begin{enumerate}
	\item update $\phi$ and $\bm{\rho}$ based on methods for different correlation structures discussed in Section \ref{section:2.2}.
	\item update $\hat{\bm{\beta}}_G$ with \eqref{4} using the Newton-Raphson method.
	\end{enumerate}
	\item Given $\hat{\bm{\beta}}_G$ or $\hat{\bm{\beta}}_Q$, update $F_{\mathfrak{p}}(\cdot)$ using \eqref{FP}.
	\item Repeat Steps 2 and 3 until convergence.
	\end{enumerate}
	
	If one is interested in obtaining the variances of the estimates of $\phi$, $\bm\rho$, and $F_{\mathfrak{p}}(t)$, the bootstrap method \citep{Dikta2021BOOK} for clustered data can be considered. That is, we can sample clusters from the original dataset with replacement to obtain $R$ bootstrap samples, and then apply the proposed method to each bootstrap sample. The variances of the estimates can be estimated with the sample variances of the estimates from the $R$ bootstrap samples.
	
	\section{Simulation studies} \label{section:3}
	
	\subsection{Data generation and parameter setting}
	We conduct extensive simulation studies to evaluate the performance of the proposed GEE and QIF methods. Because the PTC model \eqref{1} is improper, i.e., the survival function tends to a positive constant when the time approaches infinity, the common method of generating random variables is no longer valid. Instead, we generate the clustered survival data based on the relationship between the PTC model and the MC model.
	We have
    \begin{equation}
        \begin{aligned}[b]
            S(t;\bm{X}_{ij}) = &\exp (-e^{\bm{\beta}^T \bm{X}_{ij}}F_{\mathfrak{p}}(t) ) \\
            =&\exp(-e^{\bm{\beta}^T \bm{X}_{ij}}) + \{ 1 - \exp(-e^{\bm{\beta}^T \bm{X}_{ij}}) \} \frac{ \exp (-e^{\bm{\beta}^T \bm{X}_{ij}}F_{\mathfrak{p}}(t) ) -\exp(-e^{\bm{\beta}^T \bm{X}_{ij}}) }{1 - \exp(-e^{\bm{\beta}^T \bm{X}_{ij}}) }  \\
            =& 1- \pi(\bm{X}_{ij}) + \pi(\bm{X}_{ij}) S_{u}(t;\bm{X}_{ij}) , \label{13}
        \end{aligned}
    \end{equation}
    where
    \begin{equation}
        \pi(\bm{X}_{ij}) = P(Y_{ij} = 1;\bm{X}_{ij}) = 1 - \exp(-e^{\bm{\beta}^T \bm{X}_{ij}})   \label{14}
    \end{equation}
    and
    \begin{equation}
        S_{u}(t;\bm{X}_{ij}) = \dfrac{ \exp (-e^{\bm{\beta}^T \bm{X}_{ij}}F_{\mathfrak{p}}(t) ) -\exp(-e^{\bm{\beta}^T \bm{X}_{ij}}) }{1 - \exp(-e^{\bm{\beta}^T \bm{X}_{ij}}) }.      \label{15}
    \end{equation}
	Therefore, generating clustered survival data from the marginal PTC model \eqref{1} is transformed to generate clustered survival data from the marginal MC model \eqref{13} with the incidence component \eqref{14} and the latency component \eqref{15}.

     For the $i$th cluster ($i=1,2,\cdots,K$), we generate the correlated failure times of uncured individuals based on a two-step procedure.
     We first simulate multivariate standard normal random variables ($Z^*_{i1},Z^*_{i2},\cdots,Z^*_{in_i}$) following ${\bm{N}}({\bf{0}},\Sigma_i^*)$ where $\Sigma_i^*=(\tau_{ijj'})_{n_i\times n_i}$  with $\tau_{ijj'}=\tau$ if $j\neq j'$ and 1 otherwise for exchangeable structure or  $\tau_{ijj'}=\tau^{\left|j-j'\right|}$ for AR(1) structure. 
     Then we use the probability integral transform, i.e., $\tilde{T}_{ij}=F^{-1}_u(\Phi(Z^*_{ij});\bm{X}_{ij})$ to convert the marginal normal random variables to the appropriate failure times following the model \eqref{15}. Here $\Phi(\cdot)$ is the cumulative distribution function of standard normal distribution and $F_u(t;\bm{X}_{ij})=1-S_u(t;\bm{X}_{ij})$.

    To generate the correlated cure statuses $Y_{ij}$ and $Y_{ij'}$ in the $i$th cluster,
    we adopt the method proposed by \citet{Emrich1991AS} based on \eqref{14}.
    Specifically, we draw a sample $\bm{V}_i=(V_{i1}, \cdots, V_{i n_{i}})^T$ from a multivariate normal distribution $N(\bm{h}_i,\Sigma_i^{V})$
    and then dichotomize each ${V}_{ij}$ with threshold 0 to obtain $\bm{Y}_i=(Y_{i1},\cdots,Y_{in_i})^T$, i.e.,
    $Y_{ij}=1$ if $V_{ij}>0$ and 0 otherwise.
    The mean $\bm{h}_i=(h_{i1}, \cdots,h_{in_{i}})^T$
    and the correlation matrix $\Sigma_i^{V}=(\zeta_{ijj'})_{n_{i}\times n_{i}}$ are determined by
     \begin{equation}
       \begin{aligned}
           & \pi(\bm{X}_{ij})=\Phi\left(h_{ij}\right), \\
           & \eta_{ijj'}=\frac{\Phi(h_{ij},h_{ij'},\zeta_{ijj'})-\Phi(h_{ij},h_{ij'},0)}{(\pi(\bm{X}_{ij})\pi(\bm{X}_{ij'})(1-\pi(\bm{X}_{ij}))(1-\pi(\bm{X}_{ij'})))^{1/2}} \label{16}
        \end{aligned}
    \end{equation}
    such that $\bm{Y}_{i}=(Y_{i1}, \cdots, Y_{in_{i}})^T$
    has the desired moments $\bm{\pi}_{i}=(\pi(\bm{X}_{i1}),\cdots,\pi(\bm{X}_{in_{i}}))^T$ and $corr(\bm{Y}_{i})=\Sigma_i^{Y}=(\eta_{ijj'})_{n_{i}\times n_{i}}$. We set $\eta_{ijj'}=\eta$ for $j \neq j'$ and 1 otherwise for exchangeable structure and $\eta_{ijj'}=\tau^{\left|j-j'\right|}$ for AR(1) structure, respectively. 
    Here $\Phi(\cdot,\cdot;\zeta_{ijj'}) $ is the cumulative distribution function of a bivariate
    normal distribution with unit variances and correlation coefficient $\zeta_{ijj'}$.
	The $h_{ij}$ and $\zeta_{ijj'}$ can be solved from \eqref{16} uniquely, and the larger value of $\zeta_{ijj'}$ corresponds to the stronger correlation between $Y_{ij}$ and $Y_{ij'}$ given $\pi(\bm{X}_{ij})$ and $\pi(\bm{X}_{ij'})$. An R software package \texttt{mvtBinaryEP} \citep{mvtBinaryEP} is available to produce the correlated cure statuses following the procedure above.

    In the simulation study, we assume that $K = 284$ and $n_i = 9$ for clustered survival data to mimic the periodontal disease data analyzed in Section~\ref{section:4}. The true correlation structure is set as either the exchangeable or the AR(1) correlation structure characterized by $\Sigma_i^*$ and $\Sigma_i^{Y}$ jointly.
    The covariate $\bm{X}^*_{ij}=(X_{ij1},X_{ij2})^T$, where $X_{ij1}$ and $X_{ij2}$ follow the Bernoulli distribution with mean 0.5 and the uniform distribution in $(\nu,\nu+1)$, respectively. The effects of the covariates $\bm{\beta}=(\beta_0,\beta_1,\beta_2)^{T}=(-0.5,1,1)^{T}$. The censoring times are noninformative and generated from the uniform distribution $U(0,3)$. By adjusting the value of $\nu$, we consider three levels of censoring rates, i.e., 20\%, 50\%, and 90\%, corresponding to 10\%, 40\%, and 85\% cure rate respectively. We assume that the baseline cumulative distribution function $F_{\mathfrak{p}}(t) = (1-e^{-2t})I(0\leq t\leq1.5)/(1-e^{-3})$. We consider three scenarios of correlation strength, i.e. $(\eta, \tau)=(0.4, 0.8),(0.2, 0.4)$, and $(0, 0)$, which correspond to strong, weak, and no correlation among the cure statuses and the failure times of uncured patients in a cluster. We generate 1000 data sets for each setting above.

   \subsection{Simulation results}  \label{section:3.1}

  We fit each data set with the marginal PTC model \eqref{1} using the proposed GEE and QIF methods. Two working correlation structures, i.e., the exchangeable and the AR(1), are considered for each cluster. In addition, based on \eqref{8}, the inverse of the exchangeable correlation matrix can be represented by two basis matrices, i.e.,
  \begin{eqnarray*}
  \begin{aligned}
        M_{1}=\left[ \begin{array}{cccc} 1 &{}  0 &{}  \cdots &{}  0\\ 0 &{}  1 &{}  \cdots &{}  0\\ \vdots &{}  \vdots &{}  \ddots &{}  \vdots \\ 0 &{}  0 &{}  \cdots &{} 1 \end{array}\right] _{n_{i}\times n_{i}}, \quad M_{2}=\left[ \begin{array}{cccc} 0 &{}  1 &{}  \cdots &{}  1\\ 1 &{}  0 &{}  \cdots &{}  1\\ \vdots &{}  \vdots &{}  \ddots &{}  \vdots \\ 1 &{}  1 &{}  \cdots &{} 0 \end{array}\right] _{n_{i}\times n_{i}},
    \end{aligned}
    \end{eqnarray*}
	and the inverse of the AR(1) correlation matrix can be expressed by three basis matrices, i.e.,
	\begin{eqnarray*}
    \begin{aligned}
    M_{1}= & {} \left[ \begin{array}{cccc} 1 &{}  0 &{} \cdots &{}  0\\ 0 &{} 1 &{} \cdots &{}  0\\ \vdots &{} \vdots &{}  \ddots &{}  \vdots \\ 0 &{}  0 &{}  \cdots &{}  1 \end{array}\right] _{n_{i}\times n_{i}},\quad
    M_{2}=\left[ \begin{array}{cccc} 0 &{} 1 &{}  \cdots &{}  0\\ 1 &{}  0 &{}  \ddots &{}  \vdots \\ \vdots &{}  \ddots &{}  \ddots &{}  1\\ 0 &{} \cdots &{}  1 &{}  0 \end{array}\right] _{n_{i}\times n_{i}},\quad
    M_{3}= & {} \left[ \begin{array}{cccc} 1 &{}  0 &{}  \cdots &{}  0\\ 0 &{}  0 &{}  \cdots &{} 0\\ \vdots &{} \vdots &{} \ddots &{} \vdots \\ 0 &{} 0 &{} \cdots &{} 1 \end{array}\right] _{n_{i}\times n_{i}}.
    \end{aligned}
	\end{eqnarray*}
	As a comparison, we also estimate the parameters using the method proposed by \citet{Portier2017BNL} without considering the correlation structure within clusters (denoted as the NPM method). The biases, empirical variances (Var), the averages of estimated variances (Var$^{*}$), and the coverage probabilities (CP) of 95\% confidence intervals of the estimates of $\bm{\beta}=(\beta_0,\beta_1,\beta_2)^{T}$ are reported in Tables \ref{tab:1}-\ref{tab:3} with different cure rates.
	We observe in the tables that, the average estimated variances of the regression estimates from the GEE and the QIF methods are close to their empirical variances in all cases, and the 95\% confidence interval coverage rates are satisfactory and close to the nominal level. Both the GEE and the QIF methods have smaller variances than the NPM method does when the correlation exists within clusters. The QIF method has the minimum variance among the three methods, which shows that excluding the nuisance correlation parameters in the estimation procedure results in more stable estimates. The proposed methods also tend to have smaller biases in some cases.

	\begin{sidewaystable}
        \centering
		\caption{Biases, empirical variances (Var), average of estimated variances (Var$^{*}$), coverage probabilities of 95\% confidence intervals (CP) of regression coefficient estimates under cure rate 10\% and censoring rate 20\%.}
        \begin{threeparttable}
        \begin{tabular*}{\hsize}{@{}@{\extracolsep{\fill}}llrrrrrrrrrrrrrrr@{}}
            \toprule
            &       & & && \multicolumn{6}{c}{Exchangeable}                         & \multicolumn{6}{c}{AR(1)} \\
            \cmidrule(lr){6-11} \cmidrule(lr){12-17}
            \multicolumn{2}{l}{True}   & \multicolumn{3}{c}{NPM} & \multicolumn{3}{c}{GEE} & \multicolumn{3}{c}{QIF} & \multicolumn{3}{c}{GEE} & \multicolumn{3}{c}{QIF} \\
            \cmidrule(lr){3-5} \cmidrule(lr){6-8} \cmidrule(lr){9-11} \cmidrule(lr){12-14} \cmidrule(lr){15-17}
            \multicolumn{2}{l}{correlation}  & $\hat{\beta}_0$ & $\hat{\beta}_1$ & $\hat{\beta}_2$  & $\hat{\beta}_0$ & $\hat{\beta}_1$ & $\hat{\beta}_2$ & $\hat{\beta}_0$ & $\hat{\beta}_1$ & $\hat{\beta}_2$  & $\hat{\beta}_0$ & $\hat{\beta}_1$ & $\hat{\beta}_2$    & $\hat{\beta}_0$ & $\hat{\beta}_1$ & $\hat{\beta}_2$ \\
            \midrule
           \multicolumn{1}{l}{Independent}
            & Bias & 0.011  & -0.004 & -0.003 & 0.011  & -0.003 & -0.003  & 0.011  & -0.002 & -0.001 & 0.012  & -0.004  & -0.003  & 0.011  & -0.003  & -0.002     \\
            & Var  & 0.005  & 0.002  & 0.006  & 0.005  & 0.002  & 0.006   & 0.006  & 0.002  & 0.006  & 0.005  & 0.002   & 0.006   & 0.006  & 0.002   & 0.006      \\
            & Var* & 0.006  & 0.002  & 0.006  & 0.006  & 0.002  & 0.006   & 0.006  & 0.002  & 0.006  & 0.006  & 0.002   & 0.006   & 0.006  & 0.002   & 0.006      \\
            & CP   & 95.6   & 95.5   & 94.3   & 95.7   & 95.6   & 94.0    & 95.7   & 95.5   & 94.2   & 95.7   & 95.8    & 94.3    & 95.5   & 95.0    & 93.8       \\
           \multicolumn{1}{l}{Exchangeable}  & \multicolumn{12}{l}{$(\tau,\eta)=(0.4,0.8)$} &  &&&\\
            & Bias & -0.006 & 0.002  & 0.009  & -0.006 & 0.002  & 0.008  & 0.000  & 0.003  & -0.002 & -0.006  & 0.002  & 0.009  & -0.004 & 0.005  & 0.005         \\
            & Var  & 0.020  & 0.011  & 0.030  & 0.014  & 0.011  & 0.019  & 0.012  & 0.012  & 0.014  & 0.017   & 0.011  & 0.025  & 0.013  & 0.011  & 0.018         \\
            & Var* &0.020   & 0.012  & 0.028  & 0.014  & 0.011  & 0.018  & 0.012  & 0.011  & 0.013  & 0.017   & 0.011  & 0.023  & 0.014  & 0.011  & 0.016         \\
            & CP   & 94.4   & 93.6   & 93.5   & 94.9   & 94.2   & 94.7   & 95.1   & 94.0   & 95.0   & 94.7    & 94.2   & 93.4   & 95.9   & 94.0   & 94.0          \\
            & \multicolumn{12}{l}{$(\tau,\eta)=(0.2,0.4)$} &  &&&\\
            & Bias & 0.006  & 0.000  & 0.005  & 0.007  & 0.000  & 0.004  & 0.009  & 0.000  & 0.002  & 0.007  & 0.000  & 0.005  & 0.008  & 0.000  & 0.003          \\
            & Var  & 0.016  & 0.007  & 0.019  & 0.014  & 0.006  & 0.016  & 0.012  & 0.005  & 0.014  & 0.015  & 0.006  & 0.018  & 0.014  & 0.006  & 0.015          \\
            & Var* & 0.016  & 0.006  & 0.018  & 0.014  & 0.005  & 0.014  & 0.012  & 0.005  & 0.012  & 0.015  & 0.006  & 0.016  & 0.013  & 0.005  & 0.013          \\
            & CP   &  94.9  & 94.0   & 94.1   & 94.4   & 95.2   & 93.5   & 93.8   & 94.0   & 92.9   & 94.4   & 94.7   & 94.1   & 94.5   & 93.8   & 93.2           \\
            \multicolumn{1}{l}{AR(1)}  &\multicolumn{12}{l}{$(\tau,\eta)=(0.4,0.8)$} & & & &  \\
           & Bias & -0.002  & 0.002  & 0.003  & -0.001  & 0.002  & 0.001  & 0.000  & 0.001  & 0.000  & -0.002  & 0.002  & 0.003  & -0.003 & 0.003  & 0.004        \\
           & Var  & 0.013   & 0.007  & 0.020  & 0.011   & 0.007  & 0.016  & 0.010  & 0.007  & 0.015  & 0.010   & 0.007  & 0.016  & 0.009  & 0.007  & 0.013        \\
           & Var* &0.012    & 0.007  & 0.019  & 0.010   & 0.007  & 0.015  & 0.010  & 0.007  & 0.014  & 0.010   & 0.007  & 0.015  & 0.009  & 0.006  & 0.012        \\
           & CP   & 94.2    & 94.7   & 93.7   & 94.4    & 94.8   & 93.8   & 94.5   & 94.3   & 93.7   & 94.3    & 95.2   & 93.7   & 94.9   & 94.8   & 93.7         \\
            &\multicolumn{12}{l}{$(\tau,\eta)=(0.2,0.4)$}&  &&&\\
          & Bias & -0.001  & 0.002  & 0.002  & -0.001 & 0.001  & 0.001  & 0.000  & 0.002  & 0.000  & -0.001 & 0.001  & 0.002  & -0.002 & 0.001  & 0.003           \\
          & Var  & 0.012   & 0.005  & 0.012  & 0.011  & 0.004  & 0.012  & 0.010  & 0.004  & 0.011  & 0.011  & 0.004  & 0.011  & 0.010  & 0.004  & 0.010           \\
          & Var* & 0.011   & 0.005  & 0.012  & 0.010  & 0.004  & 0.011  & 0.010  & 0.004  & 0.010  & 0.010  & 0.004  & 0.010  & 0.009  & 0.004  & 0.009           \\
          & CP   &  93.8   & 93.8   & 93.9   & 93.8   & 94.1   & 93.4   & 94.2   & 92.9   & 94.0   & 93.7   & 93.2   & 93.6   & 93.6   & 93.0   & 93.3            \\
            \bottomrule
        \end{tabular*}%
		\end{threeparttable}
		\label{tab:1}%
      \end{sidewaystable}

    \begin{sidewaystable}
        \centering
		\caption{Biases, empirical variances (Var), average of estimated variances (Var$^{*}$), coverage probabilities of 95\% confidence intervals (CP) of regression coefficient estimates under cure rate 40\% and censoring rate 50\%.}
            \begin{threeparttable}
        \begin{tabular*}{\hsize}{@{}@{\extracolsep{\fill}}llrrrrrrrrrrrrrrr@{}}
            \toprule
            &       & & && \multicolumn{6}{c}{Exchangeable}                         & \multicolumn{6}{c}{AR(1)} \\
            \cmidrule(lr){6-11} \cmidrule(lr){12-17}
            \multicolumn{2}{l}{True}   & \multicolumn{3}{c}{NPM} & \multicolumn{3}{c}{GEE} & \multicolumn{3}{c}{QIF} & \multicolumn{3}{c}{GEE} & \multicolumn{3}{c}{QIF} \\
            \cmidrule(lr){3-5} \cmidrule(lr){6-8} \cmidrule(lr){9-11} \cmidrule(lr){12-14} \cmidrule(lr){15-17}
            \multicolumn{2}{l}{correlation}  & $\hat{\beta}_0$ & $\hat{\beta}_1$ & $\hat{\beta}_2$  & $\hat{\beta}_0$ & $\hat{\beta}_1$ & $\hat{\beta}_2$ & $\hat{\beta}_0$ & $\hat{\beta}_1$ & $\hat{\beta}_2$  & $\hat{\beta}_0$ & $\hat{\beta}_1$ & $\hat{\beta}_2$    & $\hat{\beta}_0$ & $\hat{\beta}_1$ & $\hat{\beta}_2$ \\
            \midrule
      \multicolumn{1}{l}{Independent}
      & Bias &  0.006 & -0.004 & -0.002 & 0.006  & -0.004 & -0.001 & 0.004 & -0.001 & 0.002  & 0.006 & -0.004 & -0.001 & 0.004 & -0.001 & 0.001   \\
      & Var  & 0.002  & 0.004  & 0.010  & 0.002  & 0.004  & 0.010  & 0.003 & 0.004  & 0.010  & 0.002 & 0.004  & 0.010  & 0.003 & 0.004  & 0.010   \\
      & Var* & 0.003  & 0.004  & 0.010  & 0.003  & 0.004  & 0.010  & 0.003 & 0.003  & 0.010  & 0.003 & 0.004  & 0.010  & 0.003 & 0.003  & 0.010    \\
      & CP   & 94.9   & 94.9   & 94.2   & 94.8   & 95.6   & 94.6   & 95.0  & 95.0   & 94.2   & 95.0  & 95.7   & 94.7   & 95.1  & 95.1   & 94.0    \\
      \multicolumn{1}{l}{Exchangeable}  & \multicolumn{12}{l}{$(\tau,\eta)=(0.4,0.8)$} &  &&&\\
      & Bias & 0.005  & 0.001  & 0.006  & 0.005  & 0.001  & 0.004  & -0.003 & 0.004  & 0.000  & 0.006  & 0.001  & 0.007  & -0.001 & 0.004  & 0.006 \\
      & Var  & 0.009  & 0.012  & 0.036  & 0.007  & 0.010  & 0.026  & 0.007  & 0.009  & 0.022  & 0.008  & 0.011  & 0.032  & 0.008  & 0.010  & 0.026 \\
      & Var* & 0.009  & 0.013  & 0.035  & 0.008  & 0.010  & 0.026  & 0.007  & 0.009  & 0.022  & 0.009  & 0.012  & 0.031  & 0.008  & 0.010  & 0.025   \\
      & CP   & 95.4   & 95.5   & 94.2   & 95.9   & 96.6   & 95.6   & 95.1   & 96.3   & 94.8   & 96.1   & 96.4   & 95.4   & 94.6   & 95.4   & 95.0  \\
      & \multicolumn{12}{l}{$(\tau,\eta)=(0.2,0.4)$} &  &&&\\
      & Bias & 0.006  & -0.001 & -0.001 & 0.006  & -0.001 & 0.000  & -0.001 & 0.003  & 0.002  & 0.006  & -0.001 & 0.000  & 0.001  & 0.002  & 0.001 \\
      & Var  & 0.007  & 0.008  & 0.030  & 0.006  & 0.006  & 0.024  & 0.006  & 0.006  & 0.020  & 0.006  & 0.007  & 0.028  & 0.006  & 0.006  & 0.023 \\
      & Var* & 0.007  & 0.008  & 0.027  & 0.006  & 0.006  & 0.022  & 0.005  & 0.005  & 0.019  & 0.006  & 0.008  & 0.025  & 0.006  & 0.006  & 0.021 \\
      & CP   & 94.5   & 95.0   & 93.3   & 95.0   & 94.9   & 94.6   & 94.0   & 94.4   & 95.5   & 95.4   & 95.1   & 94.0   & 94.6   & 93.2   & 94.7  \\
      \multicolumn{1}{l}{AR(1)}
      &\multicolumn{12}{l}{$(\tau,\eta)=(0.4,0.8)$} & & &  &  \\
      & Bias & 0.003  & 0.002  & 0.001  & 0.004  & 0.001  & 0.002  & 0.000  & 0.004  & 0.006  & 0.004  & 0.001  & 0.003  & 0.000  & 0.003  & 0.004 \\
      & Var  & 0.006  & 0.010  & 0.023  & 0.006  & 0.009  & 0.021  & 0.006  & 0.009  & 0.021  & 0.006  & 0.009  & 0.020  & 0.006  & 0.009  & 0.017  \\
      & Var* & 0.007  & 0.009  & 0.024  & 0.006  & 0.009  & 0.021  & 0.006  & 0.009  & 0.019  & 0.006  & 0.009  & 0.020  & 0.006  & 0.008  & 0.017 \\
      & CP   & 95.5   & 94.1   & 94.6   & 95.6   & 94.4   & 94.5   & 95.1   & 94.1   & 93.9   & 95.5   & 95.2   & 94.9   & 95.9   & 94.5   & 94.8  \\
      &\multicolumn{12}{l}{$(\tau,\eta)=(0.2,0.4)$}&  &&&\\
      & Bias & 0.001  & -0.003 & -0.003 & 0.002  & -0.004 & -0.003 & -0.002 & -0.002 & 0.000  & 0.001  & -0.004 & -0.005 & -0.002 & -0.002 & -0.003 \\
      & Var  & 0.004  & 0.005  & 0.015  & 0.004  & 0.005  & 0.015  & 0.004  & 0.005  & 0.014  & 0.004  & 0.005  & 0.015  & 0.004  & 0.004  & 0.014  \\
      & Var* & 0.004  & 0.005  & 0.016  & 0.004  & 0.005  & 0.015  & 0.004  & 0.004  & 0.014  & 0.004  & 0.005  & 0.015  & 0.003  & 0.004  & 0.014  \\
      & CP   & 93.9   & 94.4   & 93.7   & 94.1   & 93.9   & 94.5   & 94.2   & 93.3   & 94.8   & 94.1   & 94.3   & 94.6   & 93.9   & 94.4   & 94.5   \\
            \bottomrule
        \end{tabular*}%
    \end{threeparttable}
        \label{tab:2}%
    \end{sidewaystable}

    \begin{sidewaystable}
        \centering
		\caption{Biases, empirical variances (Var), average of estimated variances (Var$^{*}$), coverage probabilities of 95\% confidence intervals (CP) of regression coefficient estimates under cure rate 85\% and censoring rate 90\%.}
            \begin{threeparttable}
        \begin{tabular*}{\hsize}{@{}@{\extracolsep{\fill}}llrrrrrrrrrrrrrrr@{}}
            \toprule
            &       & & && \multicolumn{6}{c}{Exchangeable}                         & \multicolumn{6}{c}{AR(1)} \\
            \cmidrule(lr){6-11} \cmidrule(lr){12-17}
            \multicolumn{2}{l}{True}   & \multicolumn{3}{c}{NPM} & \multicolumn{3}{c}{GEE} & \multicolumn{3}{c}{QIF} & \multicolumn{3}{c}{GEE} & \multicolumn{3}{c}{QIF} \\
            \cmidrule(lr){3-5} \cmidrule(lr){6-8} \cmidrule(lr){9-11} \cmidrule(lr){12-14} \cmidrule(lr){15-17}
            \multicolumn{2}{l}{correlation}  & $\hat{\beta}_0$ & $\hat{\beta}_1$ & $\hat{\beta}_2$  & $\hat{\beta}_0$ & $\hat{\beta}_1$ & $\hat{\beta}_2$ & $\hat{\beta}_0$ & $\hat{\beta}_1$ & $\hat{\beta}_2$  & $\hat{\beta}_0$ & $\hat{\beta}_1$ & $\hat{\beta}_2$    & $\hat{\beta}_0$ & $\hat{\beta}_1$ & $\hat{\beta}_2$ \\
            \midrule
        \multicolumn{1}{l}{Independent}
        & Bias & -0.014  & -0.006 & -0.005 & -0.013 & -0.006 & -0.005 & -0.009 & 0.011  & 0.008  & -0.013 & -0.006 & -0.005  & -0.008 & 0.009  & 0.008 \\
        & Var  & 0.221   & 0.018  & 0.046  & 0.221  & 0.018  & 0.046  & 0.229  & 0.020  & 0.048  & 0.221  & 0.018  & 0.046   & 0.231  & 0.020  & 0.048 \\
        & Var* & 0.231   & 0.019  & 0.048  & 0.228  & 0.019  & 0.048  & 0.229  & 0.019  & 0.048  & 0.228  & 0.019  & 0.048   & 0.229  & 0.019  & 0.048 \\
        & CP   & 94.8    & 95.4   & 94.5   & 95.3   & 95.3   & 95.1   & 95.0   & 95.1   & 94.6   & 95.3   & 95.7   & 95.1    & 95.1   & 95.1   & 94.6 \\
        \multicolumn{1}{l}{Exchangeable}  &
		\multicolumn{12}{l}{$(\tau,\eta)=(0.4,0.8)$} &  &&&\\
        & Bias & -0.041 & -0.004 & -0.016 & -0.033  & -0.004 & -0.013 & -0.022 & 0.022  & 0.021  & -0.040 & -0.004 & -0.016  & -0.037 & 0.021  & 0.011  \\
        & Var  & 0.501  & 0.043  & 0.127  & 0.367   & 0.028  & 0.095  & 0.319  & 0.027  & 0.084  & 0.445  & 0.036  & 0.113   & 0.371  & 0.030  & 0.098 \\
        & Var* & 0.506  & 0.044  & 0.128  & 0.361   & 0.028  & 0.092  & 0.312  & 0.024  & 0.080  & 0.437  & 0.036  & 0.111   & 0.361  & 0.029  & 0.092 \\
        & CP   & 93.7   & 94.6   & 94.6   & 94.5    & 95.5   & 94.7   & 94.3   & 93.7   & 94.3   & 94.1   & 95.4   & 95.3    & 93.5   & 94.1   & 92.9  \\
        & \multicolumn{12}{l}{$(\tau,\eta)=(0.2,0.4)$} &  &&&\\
        & Bias & 0.005  & -0.005 & 0.004  & 0.000  & -0.006 & 0.001  & -0.021  & 0.020  & 0.015  & 0.003  & -0.005 & 0.003 & -0.010 & 0.016  & 0.018   \\
        & Var  & 0.426  & 0.034  & 0.107  & 0.356  & 0.028  & 0.091  & 0.317   & 0.027  & 0.083  & 0.406  & 0.031  & 0.102 & 0.361  & 0.029  & 0.092 \\
        & Var* & 0.412  & 0.036  & 0.104  & 0.342  & 0.028  & 0.087  & 0.297   & 0.024  & 0.076  & 0.386  & 0.033  & 0.098 & 0.333  & 0.027  & 0.084\\
        & CP   & 94.4   & 95.0   & 94.1   & 94.7   & 95.0   & 94.4   & 93.5    & 93.6   & 93.7   & 95.1   & 95.2   & 94.7  & 93.7   & 93.9   & 93.5 \\
        \multicolumn{1}{l}{AR(1)}  &\multicolumn{12}{l}{$(\tau,\eta)=(0.4,0.8)$} & &  & &  \\
        & Bias & -0.006 & -0.002 & -0.001 & -0.007 & -0.005 & -0.002 & -0.012 & 0.012  & 0.011  & -0.003 & -0.007  & -0.001 & -0.006 & 0.004  & 0.010 \\
        & Var  & 0.269  & 0.024  & 0.068  & 0.247  & 0.022  & 0.062  & 0.249  & 0.024  & 0.063  & 0.246  & 0.021   & 0.062  & 0.227  & 0.021  & 0.058\\
        & Var* & 0.275  & 0.025  & 0.069  & 0.254  & 0.022  & 0.064  & 0.246  & 0.022  & 0.062  & 0.248  & 0.022   & 0.063  & 0.229  & 0.020  & 0.058 \\
        & CP   & 94.2   & 94.7   & 94.5   & 95.0   & 94.8   & 94.9   & 95.0   & 93.5   & 94.3   & 94.5   & 94.7    & 94.5   & 94.8   & 94.4   & 94.9 \\
        &\multicolumn{12}{l}{$(\tau,\eta)=(0.2,0.4)$}&  &&&\\
        & Bias & -0.025 & -0.003 & -0.012 & -0.025 & -0.005 & -0.012 & -0.034 & 0.014  & -0.002 & -0.022 & -0.006  & -0.011 & -0.033 & 0.010  & -0.004   \\
        & Var  & 0.236  & 0.021  & 0.058  & 0.226  & 0.020  & 0.056  & 0.226  & 0.022  & 0.055  & 0.225  & 0.020   & 0.056  & 0.214  & 0.021  & 0.053 \\
        & Var* & 0.236  & 0.022  & 0.059  & 0.225  & 0.021  & 0.057  & 0.221  & 0.020  & 0.056  & 0.224  & 0.020   & 0.056  & 0.213  & 0.019  & 0.054 \\
        & CP   & 94.0   & 95.0   & 94.8   & 95.1   & 95.7   & 94.9   & 94.9   & 94.8   & 95.3   & 94.8   & 95.6    & 94.8   & 95.2   & 95.0   & 95.1  \\
            \bottomrule
        \end{tabular*}%
    \end{threeparttable}
        \label{tab:3}%
    \end{sidewaystable}
	
	To evaluate the efficiency and the optimality of the proposed methods, we calculate the ratios of the mean squared errors (MSE) of $\hat{\bm\beta}_G$ and $\hat{\bm\beta}_Q$ under the exchangeable correlation  structure versus those from the NPM method and the same ratios under the AR(1) correlation structure. The results are presented in Table \ref{tab:5}. They show that almost all values are less than 1 when the correlation exists, indicating no matter whether the working structure is correctly specified, both the GEE and the QIF methods improve the estimation efficiency by modeling the correlation explicitly.
	In addition, 
	under the same working correlation structure, the ratios related to $\hat{\bm\beta}_Q$ are generally less than those related to $\hat{\bm\beta}_G$,
	showing that the QIF method achieves higher efficiency than the GEE method.
	The efficiency gain is more substantial when the correlation is stronger, the censoring is lighter, and the working correlation structure is closer to the true correlation structure. For example, when the true correlation structure is set to the exchangeable correlation structure with $(\tau,\eta)=(0.4,0.8)$ and the censoring rate is 20\%, the relative efficiency of the estimates of $\beta_2$ in $\hat{\bm\beta}_G$ and $\hat{\bm\beta}_Q$ can be as low as 0.623 and 0.450, respectively. With the same correlation strength and cure rate aforementioned, when the true correlation structure is set to AR(1), the relative efficiency of the estimates of $\beta_2$ in $\hat{\bm\beta}_G$ and $\hat{\bm\beta}_Q$ can be as low as 0.781 and 0.649 respectively.
	All relative efficiencies are approximately equal to 1 when there is no correlation within clusters, and the three methods are comparable.

	\begin{table}[htbp]
        \centering
        \caption{Relative efficiency of the estimates from the GEE and QIF methods vs the NPM method.  }
        \begin{threeparttable}
		\begin{tabular}{lrrrrrrrrrrrr}
            \toprule
			&\multicolumn{6}{c}{Exchangeable working correlation}&\multicolumn{6}{c}{AR(1) working correlation}\\ \cmidrule(lr){2-7}  \cmidrule(lr){8-13}
			\multicolumn{1}{l}{True} &     \multicolumn{3}{c}{$\hat{\bm\beta}_G$}                   &    \multicolumn{3}{c}{$\hat{\bm\beta}_Q$} &    \multicolumn{3}{c}{$\hat{\bm\beta}_G$}  &    \multicolumn{3}{c}{$\hat{\bm\beta}_Q$}            \\
            \cmidrule(lr){2-4}  \cmidrule(lr){5-7}  \cmidrule(lr){8-10}  \cmidrule(lr){11-13}
            \multicolumn{1}{l}{correlation} & ${\beta}_0$& ${\beta}_1$& ${\beta}_2$
            & ${\beta}_0$& ${\beta}_1$& ${\beta}_2$ & ${\beta}_0$& ${\beta}_1$& ${\beta}_2$ & ${\beta}_0$& ${\beta}_1$& ${\beta}_2$ \\
            \midrule
            &\multicolumn{12}{c}{cure rate 10\%, censoring rate 20\%}  \\
            \specialrule{0em}{1pt}{1pt}
            \multicolumn{1}{l}{No correlation}                           & 1.003 & 1.005 & 1.005 & 1.020 & 1.027 & 1.011 & 1.002 & 1.004 & 1.000 & 1.021 & 1.023 & 1.015  \\
            \multicolumn{13}{l}{Exchangeable}  \\
            \multicolumn{1}{l}{- Strong}                                 & 0.717 & 0.992 & 0.623 & 0.583 & 1.028 & 0.450 & 0.880 & 0.999 & 0.836 & 0.675 & 1.016 & 0.577\\
            \multicolumn{1}{l}{- Weak}                                   & 0.889 & 0.866 & 0.856 & 0.778 & 0.793 & 0.718 & 0.964 & 0.939 & 0.951 & 0.848 & 0.846 & 0.811 \\
            \multicolumn{13}{l}{AR(1)}  \\
            \multicolumn{1}{l}{- Strong}                                 & 0.863 & 1.005 & 0.812 & 0.812 & 1.026 & 0.748 & 0.831 & 0.968 & 0.781 & 0.707 & 0.958 & 0.649\\
            \multicolumn{1}{l}{- Weak}                                   & 0.946 & 0.956 & 0.936 & 0.888 & 0.949 & 0.880 & 0.935 & 0.955 & 0.923 & 0.857 & 0.895 & 0.842  \\
            \specialrule{0em}{1pt}{1pt}
            &\multicolumn{12}{c}{cure rate 40\%, censoring rate 50\%}  \\
            \specialrule{0em}{1pt}{1pt}
            \multicolumn{1}{l}{No correlation}                           & 1.000 & 1.002 & 1.001 & 1.018 & 1.018 & 1.005 & 1.000 & 1.001 & 1.001 & 1.016 & 1.009 & 1.011 \\
            \multicolumn{13}{l}{Exchangeable}  \\
            \multicolumn{1}{l}{- Strong}                                 & 0.865 & 0.780 & 0.716 & 0.848 & 0.706 & 0.615 & 0.950 & 0.911 & 0.881 & 0.893 & 0.810 & 0.734  \\
            \multicolumn{1}{l}{- Weak}                                   & 0.892 & 0.747 & 0.807 & 0.849 & 0.666 & 0.669 & 0.959 & 0.899 & 0.926 & 0.889 & 0.756 & 0.766 \\
            \multicolumn{13}{l}{AR(1)}  \\
            \multicolumn{1}{l}{- Strong}                                 & 0.989 & 0.979 & 0.913 & 0.982 & 0.993 & 0.897 & 0.945 & 0.945 & 0.858 & 0.929 & 0.941 & 0.756 \\
            \multicolumn{1}{l}{- Weak}                                   & 0.976 & 0.952 & 0.954 & 0.981 & 0.954 & 0.922 & 0.972 & 0.938 & 0.951 & 0.943 & 0.889 & 0.896 \\
            \specialrule{0em}{1pt}{1pt}
            &\multicolumn{12}{c}{cure rate 85\%, censoring rate 90\%}  \\
            \specialrule{0em}{1pt}{1pt}
            \multicolumn{1}{l}{No correlation}                           & 1.000 & 1.004 & 1.001 & 1.036 & 1.094 & 1.043 & 1.002 & 1.001 & 1.003 & 1.045 & 1.086 & 1.047  \\
            \multicolumn{13}{l}{Exchangeable}  \\
            \multicolumn{1}{l}{- Strong}                                 & 0.732 & 0.652 & 0.751 & 0.636 & 0.642 & 0.668 & 0.887 & 0.833 & 0.892 & 0.741 & 0.722 & 0.770 \\
            \multicolumn{1}{l}{- Weak}                                   & 0.836 & 0.813 & 0.848 & 0.745 & 0.803 & 0.776 & 0.952 & 0.923 & 0.958 & 0.846 & 0.856 & 0.863 \\
            \multicolumn{13}{l}{AR(1)}  \\
            \multicolumn{1}{l}{- Strong}                                 & 0.918 & 0.893 & 0.916 & 0.927 & 0.974 & 0.935 & 0.913 & 0.875 & 0.910 & 0.845 & 0.856 & 0.857\\
            \multicolumn{1}{l}{- Weak}                                   & 0.957 & 0.958 & 0.954 & 0.960 & 1.040 & 0.947 & 0.953 & 0.949 & 0.956 & 0.909 & 0.968 & 0.903 \\
            \bottomrule
        \end{tabular}%

        \end{threeparttable}
            \label{tab:5}%
    \end{table}%

  We also examined other ratios of MSEs, such as the ratios between $\hat{\bm{\beta}}_{G}$ under the exchangeable working correlation and $\hat{\bm{\beta}}_{G}$ under the AR(1) working correlation (referred to as RE$_1$), between $\hat{\bm{\beta}}_{Q}$ under the exchangeable working correlation and $\hat{\bm{\beta}}_{Q}$ under the AR(1) working correlation (referred to as RE$_2$), between $\hat{\bm{\beta}}_{Q}$ under the exchangeable working correlation and $\hat{\bm{\beta}}_{G}$ under the AR(1) working correlation (referred to as RE$_3$), and between $\hat{\bm{\beta}}_{Q}$ under the AR(1) working correlation and $\hat{\bm{\beta}}_{G}$ under the exchangeable working correlation (referred to as RE$_4$),
  and report them in Table \ref{tab:6}.
  The results show that both $\hat{\bm{\beta}}_{G}$ and $\hat{\bm{\beta}}_{Q}$ with a correct working structure have a higher efficiency compared with those with a misspecified working structure when the correlation exists within clusters. For example, when the true correlation structure is exchangeable with $(\tau,\eta)=(0.4,0.8)$ and the censoring rate is 90\%, the RE$_1$ and RE$_2$ of $\beta_1$ are 0.783 and 0.889, respectively. When the true correlation structure is AR(1), under the same correlation strength and censoring rate, the RE$_1$ and RE$_2$ of $\beta_1$ increase to 1.021 and 1.139, separately. Table~\ref{tab:6} also shows that $\hat{\bm{\beta}}_{Q}$ with a correctly specified working structure is more efficient than $\hat{\bm{\beta}}_{G}$ with a misspecified working correlation. For example, RE$_3$ of $\beta_0$ is 0.662 with the true exchangeable correlation and RE$_4$ of $\beta_0$ is 0.819 with the true AR(1) correlation, both under $(\tau,\eta)=(0.4,0.8)$ and 20\% censoring rate.
  In addition, the MSE of $\hat{\bm{\beta}}_{Q}$ with a misspecified working correlation structure is comparable with that of $\hat{\bm{\beta}}_{G}$ with correct working correlation structure since both RE$_3$ and RE$_4$ are close to 1 in these cases. In other words, $\hat{\bm{\beta}}_{Q}$ shows robustness to the misspecified working structure compared with $\hat{\bm{\beta}}_{G}$. For example, RE$_3$ of $\hat{\beta}_0$ with the true AR(1) working correlation structure is 1.008, and RE$_4$ of $\hat{\beta}_0$ with the true exchangeable working correlation structure is 0.996, both under $(\tau,\eta)=(0.2,0.4)$ and 50\% censoring rate.

   \begin{table}[htbp]
         \centering
         \caption{Relative efficiency of the estimates between the two methods and between different working correlation structures.  }
         \begin{threeparttable}
 		\begin{tabular}{lrrrrrrrrrrrr}
             \toprule
             \multicolumn{1}{l}{True}        &  \multicolumn{3}{c}{RE$_1$}   &  \multicolumn{3}{c}{RE$_2$} &  \multicolumn{3}{c}{RE$_3$}  &  \multicolumn{3}{c}{RE$_4$}            \\
             \cmidrule(lr){2-4}  \cmidrule(lr){5-7}  \cmidrule(lr){8-10}  \cmidrule(lr){11-13}
             \multicolumn{1}{l}{correlation} & ${\beta}_0$& ${\beta}_1$& ${\beta}_2$
             & ${\beta}_0$& ${\beta}_1$& ${\beta}_2$ & ${\beta}_0$& ${\beta}_1$& ${\beta}_2$ & ${\beta}_0$& ${\beta}_1$& ${\beta}_2$ \\
             \midrule
             &\multicolumn{12}{c}{cure rate 10\%, censoring rate 20\%}  \\
             \specialrule{0em}{1pt}{1pt}
             \multicolumn{1}{l}{No correlation}                            & 1.001 & 1.001 & 1.005 & 0.998 & 1.004 & 0.996 & 1.018 & 1.023 & 1.011 & 1.018 & 1.018 & 1.010   \\
             \multicolumn{13}{l}{Exchangeable}  \\                        
             \multicolumn{1}{l}{- Strong}                                  & 0.814 & 0.994 & 0.745 & 0.863 & 1.012 & 0.780 & 0.662 & 1.030 & 0.539 & 0.941 & 1.024 & 0.927 \\
             \multicolumn{1}{l}{- Weak}                                    & 0.923 & 0.922 & 0.900 & 0.917 & 0.937 & 0.886 & 0.807 & 0.844 & 0.755 & 0.954 & 0.978 & 0.947 \\
             \multicolumn{13}{l}{AR(1)}  \\                               
             \multicolumn{1}{l}{- Strong}                                  & 1.038 & 1.038 & 1.039 & 1.148 & 1.071 & 1.153 & 0.977 & 1.060 & 0.957 & 0.819 & 0.953 & 0.799 \\
             \multicolumn{1}{l}{- Weak}                                    & 1.011 & 1.002 & 1.013 & 1.036 & 1.061 & 1.046 & 0.950 & 0.994 & 0.954 & 0.906 & 0.936 & 0.899 \\
             \specialrule{0em}{1pt}{1pt}                                  
             &\multicolumn{12}{c}{cure rate 40\%, censoring rate 50\%}  \\ 
             \specialrule{0em}{1pt}{1pt}                                  
             \multicolumn{1}{l}{No correlation}                            & 1.000 & 1.001 & 1.001 & 1.002 & 1.008 & 0.994 & 1.018 & 1.017 & 1.004 & 1.016 & 1.008 & 1.010   \\
             \multicolumn{13}{l}{Exchangeable}  \\                        
             \multicolumn{1}{l}{- Strong}                                  & 0.910 & 0.857 & 0.813 & 0.949 & 0.871 & 0.837 & 0.893 & 0.775 & 0.698 & 1.033 & 1.038 & 1.026  \\
             \multicolumn{1}{l}{- Weak}                                    & 0.929 & 0.831 & 0.872 & 0.956 & 0.881 & 0.873 & 0.885 & 0.741 & 0.722 & 0.996 & 1.013 & 0.949  \\
             \multicolumn{13}{l}{AR(1)}  \\                               
             \multicolumn{1}{l}{- Strong}                                  & 1.046 & 1.036 & 1.064 & 1.057 & 1.056 & 1.187 & 1.039 & 1.051 & 1.046 & 0.939 & 0.961 & 0.828 \\
             \multicolumn{1}{l}{- Weak}                                    & 1.004 & 1.015 & 1.003 & 1.039 & 1.073 & 1.029 & 1.008 & 1.017 & 0.970 & 0.966 & 0.934 & 0.939 \\
             \specialrule{0em}{1pt}{1pt}                                  
             &\multicolumn{12}{c}{cure rate 85\%, censoring rate 90\%}  \\ 
             \specialrule{0em}{1pt}{1pt}                                  
             \multicolumn{1}{l}{No correlation}                            & 0.999 & 1.002 & 0.999 & 0.992 & 1.007 & 0.996 & 1.034 & 1.092 & 1.040 & 1.044 & 1.082 & 1.046\\
             \multicolumn{13}{l}{Exchangeable}  \\                        
             \multicolumn{1}{l}{- Strong}                                  & 0.825 & 0.783 & 0.842 & 0.858 & 0.889 & 0.867 & 0.716 & 0.771 & 0.748 & 1.013 & 1.108 & 1.026  \\
             \multicolumn{1}{l}{- Weak}                                    & 0.879 & 0.880 & 0.885 & 0.880 & 0.938 & 0.900 & 0.782 & 0.870 & 0.810 & 1.012 & 1.053 & 1.017 \\
             \multicolumn{13}{l}{AR(1)}  \\                               
             \multicolumn{1}{l}{- Strong}                                  & 1.005 & 1.021 & 1.006 & 1.097 & 1.139 & 1.090 & 1.015 & 1.114 & 1.027 & 0.921 & 0.958 & 0.936 \\
             \multicolumn{1}{l}{- Weak}                                    & 1.004 & 1.009 & 0.998 & 1.056 & 1.074 & 1.049 & 1.008 & 1.096 & 0.991 & 0.950 & 1.011 & 0.946 \\
             \bottomrule
         \end{tabular}%
         \end{threeparttable}
             \label{tab:6}%
     \end{table}%

    The proposed GEE method also produces the estimate of $\rho$, the correlation coefficient in the working correlation matrix. Even though the estimate of $\rho$ does not correspond to the correlation measures $(\tau,\eta)$ in the data generation, Table \ref{tab:7} shows that $\hat{\rho}$ agrees well with the values of $(\tau,\eta)$ used in the data generation in the sense that when the latter increase, the former tends to increase too. When there is no correlation in clusters, $\hat{\rho}$ is very close to zero.
    To obtain the bootstrap variance of $\hat{\rho}$, we select 100 data sets randomly from the 1000 simulated data sets and choose the bootstrap sample size $R = 100$ for each scenario. From Table \ref{tab:7}, we observe that the average of 100 bootstrap variances and the empirical variances are quite close, which shows that the bootstrap variance estimator works well for calculating the variance estimate of $\hat{\rho}$ in the GEE method.

	\begin{table}[htbp]
	         \centering
	         \caption{Mean, empirical variance (Var), average of estimated variance (Var*) of $\hat{\rho}$.}
	 		\begin{tabular}{lrrrrrr}
	             \toprule
				 & \multicolumn{6}{c}{Working correlation structure}\\ \cmidrule(lr){2-7}
	             &     \multicolumn{3}{c}{Exchangeable}            &    \multicolumn{3}{c}{AR(1)}     \\
	             \cmidrule(lr){2-4}  \cmidrule(lr){5-7}
	             \multicolumn{1}{l}{True correlation}  & Mean & Var & Var*
	             &Mean & Var & Var*  \\
	             \midrule
                &\multicolumn{6}{c}{cure rate 10\%, censoring rate 20\%}  \\
             \specialrule{0em}{1pt}{1pt}
	          \multicolumn{1}{l}{No correlation}                           & 0.009 & $<0.001$ & $<0.001$ & 0.008 & $<0.001$ & $<0.001$  \\
	          \multicolumn{7}{l}{Exchangeable}  \\                       
	          \multicolumn{1}{l}{- Strong}                                 & 0.203 & 0.006    & 0.003    & 0.181 & 0.008    & 0.005 \\
	          \multicolumn{1}{l}{- Weak}                                   & 0.079 & 0.001    & 0.001    & 0.073 & 0.002    & 0.001 \\
	          \multicolumn{7}{l}{AR(1)}  \\                              
	          \multicolumn{1}{l}{- Strong}                                 & 0.155 & 0.004    & 0.002    & 0.333 & 0.015    & 0.013 \\
	          \multicolumn{1}{l}{- Weak}                                   & 0.044 & $<0.001$ & $<0.001$ & 0.111 & 0.003    & 0.002 \\
	          \specialrule{0em}{1pt}{1pt}                                 
                &\multicolumn{6}{c}{cure rate 40\%, censoring rate 50\%}  \\
	          \specialrule{0em}{1pt}{1pt}                                 
	          \multicolumn{1}{l}{No correlation}                           & 0.008 & $<0.001$ & $<0.001$ & 0.010 & $<0.001$ & $<0.001$  \\
	          \multicolumn{7}{l}{Exchangeable}  \\                       
	          \multicolumn{1}{l}{- Strong}                                 & 0.171 & 0.005    & 0.004    & 0.161 & 0.008    & 0.007  \\
	          \multicolumn{1}{l}{- Weak}                                   & 0.096 & 0.002    & 0.002    & 0.089 & 0.005    & 0.004  \\
	          \multicolumn{7}{l}{AR(1)}  \\                              
	          \multicolumn{1}{l}{- Strong}                                 & 0.104 & 0.003    & 0.002    & 0.240 & 0.014    & 0.013 \\
	          \multicolumn{1}{l}{- Weak}                                   & 0.048 & 0.001    & 0.001    & 0.121 & 0.008    & 0.006 \\
	          \specialrule{0em}{1pt}{1pt}                                 
                &\multicolumn{6}{c}{cure rate 85\%, censoring rate 90\%}  \\
	          \specialrule{0em}{1pt}{1pt}                                 
	          \multicolumn{1}{l}{No correlation}                           & 0.003 & $<0.001$ & $<0.001$ & 0.003 & $<0.001$ & $<0.001$   \\
	          \multicolumn{7}{l}{Exchangeable}  \\                       
	          \multicolumn{1}{l}{- Strong}                                 & 0.132 & 0.004    & 0.002    & 0.123 & 0.007    & 0.006  \\
	          \multicolumn{1}{l}{- Weak}                                   & 0.084 & 0.002    & 0.001    & 0.071 & 0.004    & 0.003  \\
	          \multicolumn{7}{l}{AR(1)}  \\                              
	          \multicolumn{1}{l}{- Strong}                                 & 0.059 & 0.002    & 0.001    & 0.143 & 0.008    & 0.007  \\
	          \multicolumn{1}{l}{- Weak}                                   & 0.031 & $<0.001$ & $<0.001$ & 0.091 & 0.005    & 0.004  \\
	             \bottomrule
	         \end{tabular}%
	           \label{tab:7}%
	\end{table}%
	
    \section{Periodontal disease data analysis} \label{section:4}
    We apply the proposed method to analyze periodontal disease data that were collected from patients treated at Creighton University
	School of Dentistry between August 2007 and March 2013. A subset of the original data named Tooth Loss Data is available in the R package \texttt{MST} \citep{Calhoun2018Constructing}.
    The tooth loss data includes 5336 patients and the number of teeth for each patient varies from 1 to 31. The survival time is defined as the time from trial entry to tooth loss due to dental diseases. The survival time may be censored due to trial exit. Due to the shared dietary, hygienic, and other factors, the survival times from the same patient may be correlated and they form a cluster. For the sake of simplicity, we only consider $K=284$ patients who have survival times from $n_i=9$ teeth in the data.
    The risk factors of interest include
    Mobil (1 for mobility score $\geq$ 1, and 0 otherwise),
    CAL (mean of clinical attachment loss),
    Bleeding (mean percentage of bleeding on probing), and
    Fill (percentage of filled teeth).
    The purpose of the analysis is to investigate the effects of the factors on the time to tooth loss due to dental diseases.

\begin{figure}[htbp]
        \centering
        \includegraphics[scale=0.7]{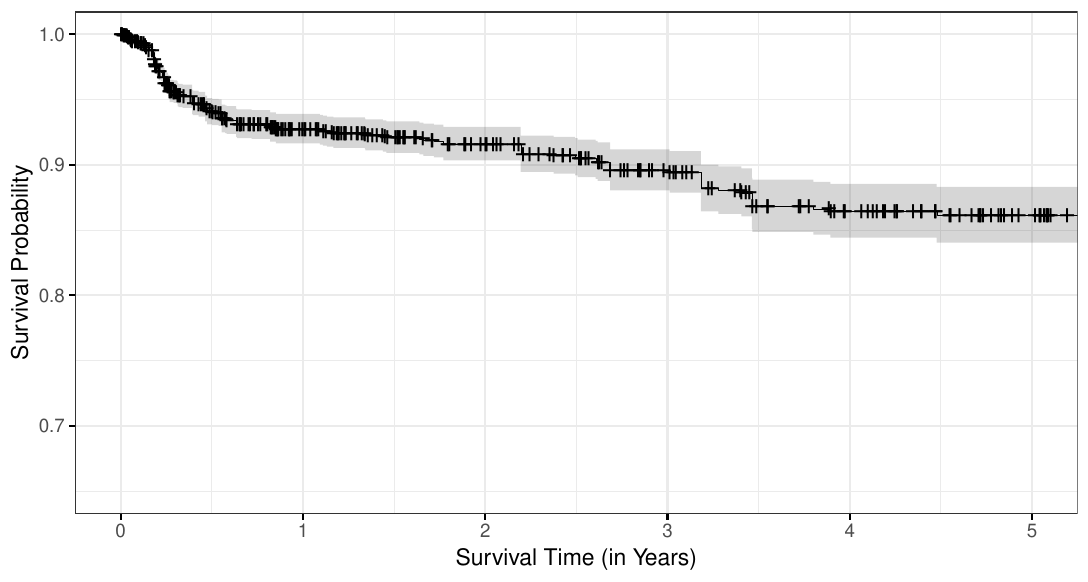}
        \caption{The Kaplan-Meier survival curve for time to tooth loss.}
        \label{fig:2}
    \end{figure}

    We plot the Kaplan-Meier survival curve of the data in Figure~\ref{fig:2}. The survival curve levels off at approximately 0.86 after 4 years of follow-up due to a large number of teeth with long-term censored times. It indicates that these teeth are likely to be immune to tooth loss due to dental diseases. This is supported by the fact that many teeth remain healthy throughout the full lifetime of a subject with well-maintained dental hygiene and they can be considered non-susceptible to dental disease progression, or ``cured'' \citep{Farewell1982BIO,Farewell1986CJS}.
    We further perform a nonparametric test \citep{Maller1992BTK} for the existence of cured or long-term survivors and obtain a $p$-value $<0.05$, showing significant evidence for a non-negligible cured fraction.
    Therefore, it is appropriate to consider a cure model for the data.

    To examine the suitability of the PTC model, following the idea of \citet{Peng2012LDA}, we consider model selection between the MC model and the PTC model based on the Box-Cox transformation cure model \citep{Yin2005CJOS}. The model is defined as follows
    $$
    \begin{aligned} S(t|\bm{X}_{ij})= \left\{ \begin{aligned}&\left\{ 1-\frac{\lambda \exp (\bm{\beta }^{T}\bm{X}_{ij})}{1+\lambda \exp (\bm{\beta }^{T}\bm{X}_{ij})} F_{\mathfrak{p}}(t)\right\} ^{1/\lambda }& 0<\lambda \leq 1 \\&\exp \{-\exp (\bm{\beta }^{T}\bm{X}_{ij})F_{\mathfrak{p}}(t)\}&\lambda =0\\ \end{aligned}   \right. \end{aligned} ,
    $$
    which is the MC model when $\lambda =1$ and the PTC model when $\lambda =0$. We select one of the two models by testing hypotheses $H_0: \lambda =0$ versus $H_1: \lambda >0$. The $p$-value of the likelihood ratio test under $H_0$ is 0.831, indicating no significant evidence against the PTC model. Therefore we consider a marginal PTC model for the data.

    We consider the proposed GEE method and QIF method as well as the NPM method
    for the data respectively. Both the exchangeable and the AR(1) working correlation structures are employed in the proposed methods. The estimates of $\bm\beta$ are summarized in Table \ref{tab:8}.
	There are two ways to interpret the estimates of $\bm\beta$: one as the log hazard ratios for overall survival time based on \eqref{1} and the other as differences in complementary log-log of cure rate based on \eqref{cure}.
    Any covariate that decreases the cure rate would also increase the overall hazard rate.
    The results in the table show that
    the GEE method and the QIF method yield smaller standard errors compared with the NPM method, and the QIF method exhibits the smallest standard errors among the three methods.
	Mobil, CAL, and Fill are significant in all three methods. That is, patients with mobility score 0 (vs $\geq 1$), a smaller mean of clinical attachment loss, or a higher percentage of filled teeth are less susceptible to tooth loss and tend to have a longer time to experience the tooth loss.
    Bleeding is significant in the QIF method with the exchangeable structure and marginally significant in the QIF method with the AR(1) structure, instead of insignificant in the NPM and the GEE methods. It implies that patients with a smaller percentage of bleeding on probing are less susceptible to tooth loss and tend to have a longer time to experience tooth loss under the proposed methods. 
    Both working structures in the GEE method reveal a positive correlation among the tooth loss times from the same patient.

	\begin{table}[htbp]
        \centering
        \caption{Estimated regression coefficients (standard errors) of the marginal PTC model with three methods}
        \begin{tabular*}{\hsize}{@{}@{\extracolsep{\fill}}lrrrrr@{}}
            \toprule
           &\multicolumn{1}{c}{NPM}&\multicolumn{2}{c}{GEE}&\multicolumn{2}{c}{QIF}\\
            \cmidrule{3-4} \cmidrule{5-6}
            \multicolumn{1}{l}{Covariates}&\multicolumn{1}{c}{} & \multicolumn{1}{c}{exchangeable} & \multicolumn{1}{c}{AR(1)} & \multicolumn{1}{c}{exchangeable} & \multicolumn{1}{c}{AR(1)} \\
            \midrule
            Intercept & -2.336 (0.472) &-2.329 (0.425) &-2.326 (0.428) & -2.257 (0.391) &-2.468 (0.397)  \\
            Mobil     &  0.886 (0.411) & 0.873 (0.379) & 0.879 (0.392) &  0.810 (0.231) & 0.633 (0.262)  \\
            CAL       &  0.264 (0.053) & 0.263 (0.056) & 0.264 (0.057) &  0.236 (0.040) & 0.253 (0.041)  \\
            Bleeding  &  0.011 (0.008) & 0.011 (0.007) & 0.011 (0.007) &  0.014 (0.006) & 0.013 (0.007)  \\
            Fill      & -1.230 (0.452) &-1.235 (0.366) &-1.238 (0.367) & -1.387 (0.361) &-1.217 (0.344)  \\
            $\rho$    & --             & 0.013 (0.015) & 0.034 (0.055) &   --           &  --            \\
            \bottomrule
        \end{tabular*}%
        \label{tab:8}%
    \end{table}%

    \section{Conclusions and discussions} \label{section:5}
    In this paper, we proposed the GEE and the QIF estimation approaches to estimate the marginal semiparametric PTC model for clustered failure time data with a cured fraction. The GEE method estimates the regression parameters in the model using an unbiased weighted estimating function with a prespecified working matrix to accommodate the correlation within clusters. The estimates from this method are shown to be consistent and asymptotically normal under regular conditions, and the variances can be consistently estimated by a sandwich variance estimator. The QIF method uses the linear combination of basis matrices to represent the inverse of the working correlation matrix in the estimating equation and can produce optimal estimates within a family of estimation equations even when the working correlation matrix is misspecified whereas the GEE estimates are not optimal in this case. The estimates from the QIF method are proved to be consistent and asymptotic normal, and the variances of the estimates are smaller than those from the GEE method.

    Our numerical study demonstrates the validity of variance estimation formulae based on the proposed methods.
    The simulation results imply that both the GEE and QIF methods improve the estimation efficiency of the regression parameters by explicitly modeling the correlation within clusters under the marginal PTC model compared to an existing method.
    The efficiency gain is more substantial when the correlation is strong, the censoring is mild, and the working structure is correctly specified.
    The estimates from the QIF method achieve higher efficiency compared with those from the GEE method under the same working structure, which shows that excluding the nuisance correlation parameters in the estimation procedure results in more stable estimates. The estimates also show robustness to the misspecified working structure compared with the estimates from the GEE method.
    When there is no correlation within clusters, the proposed two methods with the working independence assumption have similar performance as the existing method.

	The proposed marginal PTC model and the estimation methods provide a useful alternative way to model clustered survival time data with a cured fraction when the existing marginal MC model is not appropriate for the data.
    The proposed estimating procedures are easy to implement, and the algorithms are fast in convergence. We implement the proposed methods in an R package \texttt{marptc}, which is publicly available at  \href{https://github.com/Stat-FeiXiao/marptc}{https://github.com/Stat-FeiXiao/marptc}.

	\section*{Appendix A. Proof of Theorem 1}
    In this appendix, we present the details to prove Theorem \ref{thm:Theorem-1}.
    For the asymptotic normality of $\hat{\bm{\beta}}_{G}$, we require the  following regularity conditions \citep{Liang1986BTK,Yuan1998JMVA}:
	\begin{enumerate}[{A}1.]
	\item  For $\forall i =1,\cdots,K$, the expectation $E(U_i(\bm{\beta}^*))$ exists and is finite, and the variance $var(U_i(\bm{\beta}^*))$ exists and shares a common upper bound;
	\item There is a neighborhood of $\bm{\beta}^*$  on which with probability one all $\Psi_K(\bm{\beta})$  are continuously differentiable and the Jacobians $\partial\Psi_K(\bm{\beta})/\partial\bm{\beta}$ converge uniformly to a nonstochastic limit which is nonsingular at $\bm{\beta}^*$;
	\item $K^{1/2}\Psi_{{K}}(\bm{\beta}^*)\to N(0,\mathcal{V}(\bm{\beta}^*))$ in distribution where $\mathcal{V}(\bm{\beta}^*)=\lim_{{K}\to\infty}(\sum_{i=1}^{{K}}E(U_i(\bm{\beta}^*)_{i}^{{T}}U_i(\bm{\beta}^*))/{K});$
	\item $\partial\Psi_{K}(\bm{\beta}^*)/\partial\rho(\bm{\beta}^*)\to\ 0$  with probability 1;
	\item $\partial\Psi_{K}(\bm{\beta}^*)/\partial F_\mathfrak{p}(\bm{\beta}^*)\to\ 0$  with probability 1;
	\item $K^{1/2}(\hat{\phi}(\bm{\beta}^*)-\phi(\bm{\beta}^*))=O_{p}(1)$ given  $\bm{\beta}^*$;
	\item $K^{1/2}(\hat{\rho}(\bm{\beta}^*,\phi(\bm{\beta}^*),F_\mathfrak{p}(\bm{\beta}^*))-\rho(\bm{\beta}^*))=O_{p}(1)$ given $\bm{\beta}^*$, $\phi(\bm{\beta}^*)$ and $F_\mathfrak{p}(\bm{\beta}^*)$;
	\item $|\partial\hat{\rho}(\bm{\beta}^*,{\phi}(\bm{\beta}^*),F_\mathfrak{p}(\bm{\beta}^*))/\partial F_\mathfrak{p}(\bm{\beta}^*)|\leq\mathrm{G}(T,\bm{\beta}^*)$ which is $O_{p}(1)$;
	\item $|\partial\hat{\rho}(\bm{\beta}^*,\phi(\bm{\beta}^*),F_\mathfrak{p}(\bm{\beta}^*))/ \partial \phi(\bm{\beta}^*)| \leq H(T,\bm{\beta}^*)$ which is ${O}_{p}(1)$.
	\end{enumerate}

    \begin{proof}[\textbf{Proof of Theorem 1}]

	To consider the consistency of $\bm{\hat{\beta}}_{G}$ under conditions (A1) and (A2), {we need to prove
	$\sum_{i=1}^{K}E[U_i(\bm{\beta}^*)]=\bm{0}$. } {Let $H_{ij}(t)$ be the distribution function of $\tilde{T}_{ij}$ and $\mu_{ij} = \mu(X_{ij})$.} We have
	$$
	\begin{aligned}
	    E\{(U^G(\bm{\beta}^*))_{ij}|C_{ij}\}& =\int\sum_{l=1}^{n_{i}}\phi^{-1}X_{il}\mu_{il}^{1/2}\tilde{Q}_{i}(\rho)_{lj}\mu_{ij}^{-1/2}F_\mathfrak{p}(t|C_{ij})(\kappa_{ij}-\mu_{ij})dH_{ij}(t|C_{ij},\theta)  \\
	    &=M_{ij}\int X_{ij}(\delta_{ij}-F_\mathfrak{p}(t|C_{ij})\mu_{ij})d H_{ij}(t|C_{ij}) \\
	    &=M_{ij}\int\frac{\partial\mu_{ij}}{\partial\beta}\frac1{\mu_{ij}}(\delta_{ij}-F_\mathfrak{p}(t|C_{ij})\mu_{ij})d H_{ij}(t|C_{ij}) \\
	    &=M_{ij}\int\frac\partial{\partial\beta}\log[\{f_\mathfrak{p}(t|c_{ij},\theta)\mu_{ij}\}^{\delta_{ij}}\exp(-F_\mathfrak{p}(t|C_{ij})\mu_{ij})]dH_{ij}(t|C_{ij})=\mathbf{0},
	\end{aligned}
	$$
	where $M_{ij}=\{\phi^{-1}\sum_{l=1}^{n_{i}}X_{il}\mu_{il}^{1/2}\tilde{Q}_{\mathrm{i}}(\rho)_{lj}\mu_{ij}^{-1/2}/X_{ij}\}$ and $\tilde{Q}_i(\rho)={Q}_i^{-1}(\rho)$ and ${Q}_i(\rho)$ is an $n_i\times n_i$ working correlation matrix for ${i}=1,\ldots,K.$ Thus $E\{(U^G(\bm{\beta}^*))_{ij}\}=E(E\{(U^G(\bm{\beta}^*))_{ij}|C_{ij}\})=\mathbf{0}$ and $\sum_{i=1}^{K}E[U^G_i(\bm{\beta}^*)]=\bm{0}.$

	To prove the asymptotic normality of $\hat{\bm{\beta}}_{G}$,
	we let $U^{G(r)}(\cdot)$ denote the derivative of $U^G(\cdot)$ with respect to the $r$th component. Following \citet{Liang1986BTK}, we have
	{$$
	K^{1/2}(\hat{\bm{\beta}}_{G}-\bm{\beta}^*)\approx\frac{\sum\limits_{i=1}^{K}U_{i}(\bm{\beta}^*,\hat{\rho}(\bm{\beta}^*,\hat{\phi}(\bm{\beta}^*),\hat{F}_\mathfrak{p}(\bm{\beta}^*)),\hat{F}_\mathfrak{p}(\bm{\beta}^*))/K^{1/2}}{\sum\limits_{i=1}^{K}\left[-\frac{\partial}{\partial \bm{\beta}}U_{i}(\bm{\beta}^*,\hat{\rho}(\bm{\beta}^*,\hat{\phi}(\bm{\beta}^*),\hat{F}_{n}(\bm{\beta}^*)),\hat{F}_\mathfrak{p}(\bm{\beta}^*))/K\right]},
	$$}
	where
	$$
	\begin{aligned}
	    &\frac{\partial}{\partial\beta}U^G_{i}(\bm{\beta}^*,\hat{\rho}(\bm{\beta}^*,\hat{\phi}(\bm{\beta}^*),\hat{F}_\mathfrak{p}(\bm{\beta}^*)),\hat{F}_\mathfrak{p}(\bm{\beta}^*)) \\
	    &=U_{i}^{G(1)}(\bm{\beta}^*,\hat{\rho}(\bm{\beta}^*,\hat{\phi}(\bm{\beta}^*),\hat{F}_\mathfrak{p}(\bm{\beta}^*)),\hat{F}_\mathfrak{p}(\bm{\beta}^*))  \\
	    +&U_{i}^{G(2)}(\bm{\beta}^*,\hat{\rho}(\bm{\beta}^*,\hat{\phi}(\bm{\beta}^*),\hat{F}_\mathfrak{p}(\bm{\beta}^*)),\hat{F}_\mathfrak{p}(\bm{\beta}^*)) \frac{\partial\hat{\rho}(\bm{\beta}^*,\hat{\phi}(\bm{\beta}^*),\hat{F}_\mathfrak{p}(\bm{\beta}^*))}{\partial\beta} \\
	    +&U_{i}^{G(3)}(\bm{\beta}^*,\hat{\rho}(\bm{\beta}^*,\hat{\phi}(\bm{\beta}^*),\hat{F}_\mathfrak{p}(\bm{\beta}^*)),\hat{F}_\mathfrak{p}(\bm{\beta}^*)) \frac{\partial\hat{F}_\mathfrak{p}(\bm{\beta}^*)}{\partial\beta} \\
	    &=A_{i}^{*}+B_{i}^{*}C^{*}+D_{i}^{*}E^{*}.
	\end{aligned}
	$$
	By Taylor expansion around point $(\bm{\beta}^*,\rho(\bm{\beta}^*),F_\mathfrak{p}(\bm{\beta}^*)),$ we have
	$$
	\begin{aligned}
	    &\frac{\sum_{i=1}^{n}U^G_{i}(\bm{\beta}^*,\hat{\rho}(\bm{\beta}^*,\hat{\phi}(\bm{\beta}^*),\hat{F}_\mathfrak{p}(\bm{\beta}^*)),\hat{F}_\mathfrak{p}(\bm{\beta}^*))}{K^{1/2}} \\
	    =&\frac{\sum_{i=1}^{K}U^G_{i}(\bm{\beta}^*, \rho(\bm{\beta}^*),F_\mathfrak{p}(\bm{\beta}^*))}{K^{1/2}}\\
	    &+\frac{\partial/\partial\rho(\bm{\beta}^*)\sum_{i=1}^{K}U^G_{i}(\bm{\beta}^*,\rho(\bm{\beta}^*),F_\mathfrak{p}(\bm{\beta}^*))}{K^{1/2}} \times (\hat{\rho}(\bm{\beta}^*,\hat{\phi}(\bm{\beta}^*),\hat{F}_\mathfrak{p}(\bm{\beta}^*))-\rho(\bm{\beta}^*)) \\
	    &+\frac{\partial/\partial F_\mathfrak{p}(\bm{\beta}^*)\sum_{i=1}^{K}U^G_{i}(\bm{\beta}^*, \rho(\bm{\beta}^*),F_\mathfrak{p}(\bm{\beta}^*))}{K^{1/2}}(\hat{F}_\mathfrak{p}(\bm{\beta}^*)-F_\mathfrak{p}(\bm{\beta}^*))+o_{p}(1) \\
	    =&A^{**}+B^{**}C^{**}+D^{**}E^{**}+o_{p}(1),
	\end{aligned}
	$$
	where $B^{**}=o_{p}(1)$ and $D^{**}=o_{p}(1)$ by (A4) and (A5).
	$$
	\begin{aligned}
	    C^{**} &=\hat{\rho}(\bm{\beta}^*,\hat{\phi}(\bm{\beta}^*),\hat{F}_\mathfrak{p}(\bm{\beta}^*))-\hat{\rho}(\bm{\beta}^*,\hat{\phi}(\bm{\beta}^*),F_\mathfrak{p}(\bm{\beta}^*))\\
	    &+\hat{\rho}(\bm{\beta}^*,\hat{\phi}(\bm{\beta}^*),F_\mathfrak{p}(\bm{\beta}^*))-\hat{\rho}(\bm{\beta}^*,\phi(\bm{\beta}^*),F_\mathfrak{p}(\bm{\beta}^*))+\hat{\rho}(\bm{\beta}^*,\phi(\bm{\beta}^*),F_\mathfrak{p}(\bm{\beta}^*))-\rho(\bm{\beta}^*)  \\
	    &=\frac{\partial\hat{\rho}(\bm{\beta}^*,\hat{\phi}(\bm{\beta}^*),F_\mathfrak{p}^*)}{\partial F_\mathfrak{p}(\bm{\beta}^*)}(\hat{F}_\mathfrak{p}(\bm{\beta}^*)-F_\mathfrak{p}(\bm{\beta}^*))+\frac{\partial\hat{\rho}(\bm{\beta}^*,\phi^*,F_\mathfrak{p}(\bm{\beta}^*))}{\partial\phi(\bm{\beta}^*)}(\hat{\phi}(\bm{\beta}^*)-\phi(\bm{\beta}^*)) \\
	    &+\hat{\rho}(\bm{\beta}^*,\phi(\bm{\beta}^*),F_\mathfrak{p}(\bm{\beta}^*))-\rho(\bm{\beta}^*)=O_{p}(1)
	\end{aligned}
	$$
	by (A6) - (A9) and $E^{**} = O_p(1)$ by \citet{Portier2017BNL}, where $F_\mathfrak{p}^*$ is between $\hat{F}_\mathfrak{p}(\bm{\beta}^*)$ and $F_\mathfrak{p}(\bm{\beta}^*)$ and $\phi^*$ is between $\hat{\phi}(\bm{\beta}^*)$ and $\phi(\bm{\beta}^*)$.\par
	
	Therefore, $$\frac{\sum_{i=1}^{K}U^G_{i}(\bm{\beta}^*,\hat{\rho}(\bm{\beta}^*,\hat{\phi}(\bm{\beta}^*),\hat{F}_\mathfrak{p}(\bm{\beta}^*)),\hat{F}_\mathfrak{p}(\bm{\beta}^*))}{K^{1/2}}$$ is asymptotically equivalent to $A^{**}. $ From (A3), we know that the asymptotic distribution $A^{**}$ is multivariate normal distribution with zero mean and covariance matrix $\lim\limits_{K\to\infty}(\sum_{i=1}^{K}E((U(\bm{\beta}^*))_{i}^{T}(U(\bm{\beta}^*))_{i})/K).$ In addition, we have $\sum_{i=1}^{K}B_{i}^{*}/K = o_{p}(1),\ \sum_{i=1}^{K}D_{i}^{*}/K = o_{p}(1),\ C^{*}=O_{p}(1) ,\  E^{*}=O_{p}(1).$ Therefore, $$\sum_{i=1}^{K}\frac{\partial}{\partial\beta}U^G_{i}(\bm{\beta}^*,\hat{\rho}(\bm{\beta}^*,\hat{\phi}(\bm{\beta}^*),\hat{F}_\mathfrak{p}(\bm{\beta}^*)),\hat{F}_\mathfrak{p}(\bm{\beta}^*))/K$$ is
	asymptotically equivalent to which converges to $\sum_{i=1}^{K}A_{i}^{*}/K$ which converges to $\frac{1}{K}\partial U^G(\bm{\beta}^*)/\partial\beta$ as $K \rightarrow \infty$. By the Slutsky's theorem, as $K \rightarrow \infty,$
	$$ K^{1/2}(\hat{\beta}_{G}-\bm{\beta}^*)\to N(0,\Sigma_{G}(\bm{\beta}^*))\quad\text{in distribution,}$$
	where $\Sigma_{G}(\bm{\beta}^*)=\mathcal{A}^{-1}(\bm{\beta}^*)\mathcal{V}(\bm{\beta}^*)\mathcal{A}^{-T}(\bm{\beta}^*),\mathcal{A}(\bm{\beta}^*)=-\partial U^G(\bm{\beta}^*)/\partial\beta$ and $\mathcal{V}(\bm{\beta}^*)=\sum_{i=1}^K\{U^G(\bm{\beta}^*)\}_i\{U^G(\bm{\beta}^*)\}_i^T.$
	\end{proof}
   Next, we show the details of calculating $\Sigma_{G}(\bm{\beta}^*)$. Obtaining $\mathcal{V}(\bm{\beta}^*)$ is straightforward. For $\mathcal{A}(\bm{\beta}^*)$, the $(\nu,\omega)\{\omega,\nu=1,2,\ldots,p_X+1\}$ element is
	$$ \sum_{i=1}^{K}A_{i}[B_{i}W_{i}C_{i}-D_{i}W_{i}E_{i}],$$
	where
	$$
	\begin{gathered}
	   A_{i} =(X_{i1\nu},X_{i2\nu},\ldots,X_{in_{i}\nu})_{1\times n_{i}}, \\
	   B_{i} =\begin{pmatrix}B_{i11}&\cdots&B_{i1n_{i}}\\\vdots&\ddots&\vdots\\B_{in_{i}1}&\cdots&B_{in_{i}n_{i}}\end{pmatrix}_{n_{i}\times n_{i}}
	\end{gathered}
	$$
	with
	$$
	B_{imn}=\frac{1}{2\phi}(X_{im\omega}-X_{in\omega})(\mu_{im})^{1/2}(\mu_{in})^{-1/2}\tilde{Q}_{i}(\rho)_{mn}
	$$
	for $m\neq n,$ otherwise $B_{imm}=0,m=1,2,\ldots,n_{i}.$
	$$
	\begin{aligned}
	   &C_{i} =(\kappa_{i1}-\mu_{i1},\kappa_{i2}-\mu_{i2},\ldots,\kappa_{in_{i}}-\mu_{in_{i}})^{T},  \\
	   &D_{i} =\begin{pmatrix}D_{i11}&\cdots&D_{i1n_{i}}\\\vdots&\ddots&\vdots\\D_{in_{i}1}&\cdots&D_{in_{i}n_{i}}\end{pmatrix}_{n_{i}\times n_{i}}
	\end{aligned}
	$$
	with
	$$
	D_{imn}^{(\beta)}=\phi^{-1}(\mu_{im})^{1/2}(\mu_{in})^{-1/2}\tilde{Q}_{i}(\rho)_{mn}
	$$
	for $m\neq n,$ otherwise $D_{imm}=\tilde{Q}_{i}(\rho)_{mm},$
	$
	E_{i}=\left(X_{i1\omega}\mu_{i1},X_{i2\omega}\mu_{i2},\ldots,X_{in_{i}\omega}\mu_{in_{i}}\right)^{T}
	$
	and $W_{i}=\mathrm{diag}(F_\mathfrak{p}(t_{i1}),\ldots,F_\mathfrak{p}(t_{in_{l}})).$
    \section*{Appendix B. Proof of Theorem 2}
    The regularity conditions based on \citet{Hansen1982Eco} and  \citet{Niu2015CSDA} are provided as follows:
	
	\begin{enumerate}[{B}1.]
	\item     $U^Q(\bm{\beta}^*)=[\sum _{i=1}^{K}{\dot{G}}_{K}^{T}(\bm{\beta}^*)C_{K}^{-1}(\bm{\beta}^*)g_{i}(\bm{\beta}^*)]/K$  converges to 0 with probability 1;
	\item     $\partial U^Q(\bm{\beta}^*)/\partial F_\mathfrak{p}(\bm{\beta}^*) $ converges to 0 with probability 1;
	\item     $G_K(\cdot)$ is a continuously differentiable function in the parameter space;
	\item     The expectation $\textrm{E}[G_K(\bm{\beta})]$ of $G_K(\bm{\beta})$ exists and is finite with a unique zero point $\bm{\beta}^*$ in the parameter space;
	\item     When $K\rightarrow \infty$, the random matrix $C_K$ converges to the constant matrix $C_0$ with probability 1;
	\item     The first derivative of $G_K$ with respective to $\bm{\beta}$ exists and is continuous.
	\end{enumerate}

  \begin{proof}[\textbf{Proof of Theorem 2}]
    By Taylor expansion we have
    $$ \sqrt{K} (\hat{\bm{\beta}} _{Q}-\bm{\beta}^* ) \thickapprox - \dfrac {U^Q\left(  \bm{\beta}^*,{\hat{F}}_\mathfrak{p}(\bm{\beta}^*) \right) /K^{1/2}} { \frac{\partial }{\partial \beta} U^Q\left( \bm{\beta}^*,\hat{F}_\mathfrak{p} (\bm{\beta}^*) \right)/K} .$$
    Similarly, let $U^{Q(r)}(\cdot)$ denote the derivative of $U^Q(\cdot)$ with respective to the $r$th component. Taylor expansion around point $(\bm{\beta}^*, F_\mathfrak{p}(\bm{\beta}^*))$ gives
    $$ \begin{aligned} \sqrt{K}U^Q\left( \bm{\beta}^*,\hat{{F }}_\mathfrak{p}(\bm{\beta}^*)\right) =&\sqrt{K}U^Q\left( \bm{\beta}^*,{{F }}_\mathfrak{p}(\bm{\beta}^*)\right)+U^{Q(2)}\left(\bm{\beta}^*,{{F }}_\mathfrak{p}(\bm{\beta}^*)\right) \sqrt{K}\left( {\hat{F }}_\mathfrak{p}(\bm{\beta}^*)-F_\mathfrak{p}(\bm{\beta}^*)\right) +o_{p}(1)\\ =&A+BC+o_{p}(1), \end{aligned}$$
    where $B=o_{p}(1)$ by (B2)  and $E =O_{p}(1)$ by \citet{Portier2017BNL}. Therefore, $\sqrt{K}U^Q\left( \bm{\beta}^*,{\hat{F }}_\mathfrak{p}(\bm{\beta}^*)\right)$ can be asymptotically approximated by $\sqrt{K}U^Q\left( \bm{\beta}^*,{F}_\mathfrak{p}(\bm{\beta}^*)\right).$
    $$\begin{aligned} \frac{\partial }{\partial \beta}U^Q\left(\bm{\beta}^*,\hat{F }_\mathfrak{p}(\bm{\beta}^*)\right) =&\, U^{Q(1)}\left(\bm{\beta}^*,\hat{F }_\mathfrak{p}(\bm{\beta}^*)\right)+U^{Q(2)}\left(\bm{\beta}^*,\hat{{F }}_\mathfrak{p}(\bm{\beta}^*)\right) \frac{\partial {\hat{F }}_\mathfrak{p}(\bm{\beta}^*)}{\partial \bm{\beta}^*} \\ =&\, D+EE^*. \end{aligned}$$
    Similarly, $E=o_{p}(1)$ by (B2) and $E^*=O_{p}(1)$ as discussed in the Appendix A. Hence, $\partial U^Q(\bm{\beta}^*,\hat{F }_\mathfrak{p}(\bm{\beta}^*))/\partial \beta$ is asymptotically equivalent to $U^{Q(1)}\left(\bm{\beta}^*,\hat{F }_\mathfrak{p}(\bm{\beta}^*)\right)$. By a similar discussion  as \citet{Spiekerman1998JASA}, we know that $\partial U^Q(\bm{\beta}^*,\hat{F}_\mathfrak{p}(\bm{\beta}^*))/\partial \bm{\beta}$ converges to $\partial U^Q(\bm{\beta}^*)/\partial \bm{\beta}$ as $K\rightarrow \infty$. Thus,
    $$\begin{aligned} \sqrt{K}({\hat{\bm{\beta} }_{Q}}-\bm{\beta}^*)\rightarrow -\left[ \frac{\partial U^Q(\bm{\beta}^*)}{\partial \bm{\beta}}\right] ^{-1}\sqrt{K}U^Q(\bm{\beta}^*) \end{aligned}$$
    with $K\rightarrow \infty$. From \citet{Qu2000BTK},  $\partial U^Q(\bm{\beta}^*)/\partial \bm{\beta}$ an be asymptotically approximated by ${\dot{G}}_{K}^{T}(\bm{\beta}^* )C_{K}^{-1}(\bm{\beta}^*){\dot{G}}_{K}(\bm{\beta}^*)$, the following equation can be derived:
    $$\begin{aligned} \left[ \frac{\partial U^Q(\bm{\beta}^*)}{\partial \bm{\beta}}\right] ^{-1}\sqrt{K}U^Q(\bm{\beta}^*) \approx &\left[ {\dot{G}}_{K}^{T}(\bm{\beta}^*)C_{K}^{-1}(\beta){\dot{G}}_{K}(\bm{\beta}^*)\right] ^{-1}\sqrt{K}U^Q(\bm{\beta}^*)\\ =&\left[ {\dot{G}}_{K}^{T}(\bm{\beta}^*)C_{K}^{-1}(\bm{\beta}^*){\dot{G}}_{K}(\bm{\beta}^*)\right] ^{-1}\\&\times {\dot{G}}_{K}^{T}(\bm{\beta}^*)C_{K}^{-1}(\bm{\beta}^*)\left\{ \sqrt{K}G_{K}(\bm{\beta}^*)\right\} . \end{aligned}$$
    Therefore, based on the above equation we have
    \begin{equation}
        \begin{aligned} \sqrt{K}({\hat{\bm{\beta}}_Q}-\bm{\beta}^*)\rightarrow -&\left[ {\dot{G}}_{K}^{T}(\bm{\beta}^*)C_{K}^{-1}(\bm{\beta}^*){\dot{G}}_{K}(\bm{\beta}^*)\right] ^{-1} {\dot{G}}_{K}^{T}(\bm{\beta}^*)C_{K}^{-1}(\bm{\beta}^*)\left\{ \sqrt{K}G_{K}(\bm{\beta}^*)\right\} \label{20}
        \end{aligned}
    \end{equation} in probability.
    
    Since $X_{i}(i=1,\ldots ,K)$ are assumed to be independent and identically distributed in clustered failure time data, the extended scores $g_{i}(\beta )(i=1,\ldots ,K)$
    are also independent and identically distributed. From (B4), we know that $\textrm{E}[G(\bm{\beta}^*)]=0$, so the variance of $G(\bm{\beta}^*)$ is $\textrm{E}[G(\bm{\beta}^*)]^{2}$. Because $G_K$ is the first-order sample moment function of $G$,  by the central limit theorem, we have
    $$\begin{aligned} \frac{1}{K}\sum _{i=1}^{K}g_{i}(\bm{\beta}^*)\rightarrow N\left( 0,\frac{1}{K}\sum _{i=1}^{K}g_{i}(\bm{\beta}^*)g_{i}^{T}(\bm{\beta}^*)\right) \end{aligned}  $$
    in distribution. Therefore, $\sqrt{K}G_{K}(\bm{\beta}^*)$ converges in distribution to a normal distribution, that is
    $$\begin{aligned} \sqrt{K}G_{K}(\bm{\beta}^*)\rightarrow N\left( 0,\sum _{i=1}^{K}g_{i}(\bm{\beta}^*)g_{i}^{T}(\bm{\beta}^*)\right) . \end{aligned}  $$
    Put the above results into \eqref{20}, we have $\sqrt{K}(\hat{\bm{\beta}}_{Q}-\bm{\beta}^*)\rightarrow N(0,\Sigma_{Q}(\bm{\beta}^*))$ in distribution, where
    $$\begin{aligned} &  \Sigma_{Q}(\bm{\beta}^*) =\left[ -{\dot{G}}_{K}^{T}(\bm{\beta}^* )C_{K}^{-1}(\bm{\beta}^*){\dot{G}}_{K}(\bm{\beta}^*)\right] ^{-1}\Omega (\bm{\beta}^*)\left[ -{\dot{G}}_{K}^{T}(\bm{\beta}^*)C_{K}^{-1}(\bm{\beta}^*){\dot{G}}_{K}(\bm{\beta}^*)\right] ^{-T}, \\ &  \Omega (\bm{\beta}^*) = \sum _{i=1}^{K}U^Q_{i}(\bm{\beta}^*)(U^Q_{i}(\bm{\beta}^*))^{T}, \\{} & {} U^Q_{i}(\bm{\beta}^*) = {\dot{G}}_{K}^{T}(\bm{\beta}^*)C_{K}^{-1}(\bm{\beta}^*)g_{i}(\bm{\beta}^*). \end{aligned}  $$
    
    Furthermore, we have
    $$ \begin{aligned} \hat{\bm{\beta}}_Q \rightarrow N \left( \bm{\beta}^*,\frac{1}{K}\Sigma_{Q}(\bm{\beta}^*) \right) \end{aligned}  $$
    in distribution, so $ \hat{\bm{\beta}}_Q\rightarrow \bm{\beta}^*$ in distribution. We know that $ \hat{\bm{\beta}}_{Q}\rightarrow \bm{\beta}^* $ in probability can be obtained by $\hat{\bm{\beta}}_Q\rightarrow \bm{\beta}^*$ in distribution, so $\hat{\bm{\beta}}_Q$ is a consistent estimator of $\bm{\beta}^*$.
     \end{proof}
	Next, we show the details of the variance calculation of $\hat{\bm{\beta}}_Q$. Here we present the calculation for ${\dot{G}}_{K}(\bm{\beta}^* )$, as the computations for the other components of $\Sigma_{Q}(\bm{\beta}^*)$ are straightforward. Let
	$$\begin{aligned} {\dot{G}}_{K}(\bm{\beta}^* )=\left[ \begin{array}{c} {\dot{G}}_{K}^{1}(\bm{\beta}^* )\\ \vdots \\ {\dot{G}}_{K}^{m}(\bm{\beta}^* ) \end{array}\right] , \end{aligned}$$
	where ${\dot{G}}_{K}^{j}(\bm{\beta}^* )$ is a $(p_{X}+1)\times (p_{X}+1)$ dimensional matrix. Specifically, using the symbols for $A_i$, $W_i$, $C_i$, and $E_i$ as defined in Appendix A, the $(v,\omega )\left\{ v,\omega =1,2,\ldots ,p_X+1\right\}$ element in the matrix ${\dot{G}}_{K}^{j}$ is
	$$\begin{aligned} \sum _{i=1}^{K}A_{i}\left[ B_{i}^{j}W_{i}C_{i}^{j}-D_{i}^{j}W_{i}E_{i}\right] , \end{aligned} $$
	where
	$$\begin{aligned}  B_{i}^{j}=\left( \begin{array}{ccc} B_{i11}^{j} & \quad \cdots & \quad B_{i1n_{i}}^{j}\\ \vdots & \quad \ddots & \quad \vdots \\ B_{in_{i}1}^{j} & \quad \cdots & \quad B_{in_{i}n_{i}}^{j} \end{array}\right) _{n_{i}\times n_{i}} \end{aligned}  $$
	with
	$$\begin{aligned} B_{imn}^{j}=\frac{1}{2}(X_{im\omega }-X_{in\omega })(\mu _{im})^{1/2}(\mu _{in})^{-1/2}(M_{j})_{mn} \end{aligned}$$
	and
	$$ \begin{aligned} &  D_{i}^{j}=\left( \begin{array}{ccc} D_{i11}^{j} & \quad \cdots & D_{i1n_{i}}^{j}\\ \vdots & \quad \ddots & \quad \vdots \\ D_{in_{i}1}^{j} & \quad \cdots & \quad D_{in_{i}n_{i}}^{j} \end{array}\right) _{n_{i}\times n_{i}} \end{aligned} $$
	with
	$$\begin{aligned} D_{imn}^{j}=(\mu _{im})^{1/2}(\mu _{in})^{-1/2}(M_{j})_{mn}. \end{aligned}$$

	We now show that $\hat{\bm\beta}_Q$ has a smaller asymptotic variance than $\hat{\bm\beta}_G$ does in (c).
    For the sake of simplicity, we drop $\bm\beta$ in the following equations.
    Since $U^G(\bm{\beta})=K Z G_K(\bm{\beta})=Z \left( \sum _{i=1}^{K}g_{i}\right)$, where $Z=\left[ Z_{1}\cdots Z_{m}\right]$, the covariance matrix of $\hat{\bm{\beta}}_{G}$ can be written as
    $$
    \begin{aligned} \Sigma_G=\frac{1}{K}\left\{ \left[ KZ{\dot{G}}_{K}^{T}\right] ^{T}\left[ \sum _{i=1}^{K}\left( Zg_{i}\right) \left( Zg_{i}\right) ^{T}\right] ^{-1}\left[ KZ{\dot{G}}_{K}^{T}\right] \right\}^{-1}=\left[ {\dot{G}}_{K}^{T}Z^{T}\left( ZC_{K}Z^{T}\right) ^{-1}Z{\dot{G}}_{K}\right] ^{-1}. \end{aligned}
    $$
    Note that
    $$
    \begin{aligned}{} & {} \left[ \begin{array}{cc} C_{K}^{-1}-Z^{T}\left( ZC_{K}Z^{T}\right) ^{-1}Z &{} \quad 0\\ 0 &{} \quad \,ZC_{K}Z^{T} \end{array}\right]  \\= & {} \left[ \begin{array}{cc} I &{} \quad \,-Z^{T}\left( ZC_{K}Z^{T}\right) ^{-1}\\ 0 &{} \quad I \end{array}\right] \left[ \begin{array}{cc} C_{K}^{-1} &{} \quad Z^{T}\\ Z &{} \quad \,ZC_{K}Z^{T} \end{array}\right] \left[ \begin{array}{cc}  I  &{} \quad 0\\ -\left( ZC_{K}Z^{T}\right) ^{-1}Z\, &{} \quad  I  \end{array}\right]
        \\= & {} \left[ \begin{array}{cc} I &{} \quad \,-Z^{T}\left( ZC_{K}Z^{T}\right) ^{-1}\\ 0 &{} \quad I \end{array}\right] \left[ \begin{array}{c} C_{K}^{-1/2}\\ ZC_{K}^{1/2} \end{array}\right] \left[ \begin{array}{cc} C_{K}^{-1/2}&\quad C_{K}^{1/2}Z^{T}\end{array}\right] \left[ \begin{array}{cc} I &{} \quad 0\\ -\left( ZC_{K}Z^{T}\right) ^{-1}Z\, &{} \quad I \end{array}\right] \end{aligned}
    $$
    is a non-negative definite matrix. Therefore, $C_{K}^{-1}-Z^{T}\left( ZC_{K}Z^{T}\right) ^{-1}Z$ is also a non-negative definite matrix. Thus $C_{K}^{-1}\ge Z^{T}\left( ZC_{K}Z^{T}\right) ^{-1}Z$. On the other hand, the covariance matrix of $\hat{\bm{\beta}}_{Q}$ in the Theorem \ref{thm:Theorem-2} can be simplified as follows,
    $$
    \begin{aligned} \Sigma_Q=&\frac{1}{K}\left[ {\dot{G}}_{K}^{T}C_{K}^{-1}{\dot{G}}_{K}\right] ^{-1}\sum _{i=1}^{K}\left( {\dot{G}}_{K}^{T}C_{K}^{-1}g_{i}\right) \left( {\dot{G}}_{K}^{T}C_{K}^{-1}g_{i}\right) ^{T}\left[ {\dot{G}}_{K}^{T}C_{K}^{-1}{\dot{G}}_{K}\right] ^{-T}\nonumber \\ =&\frac{1}{K}\left\{ \left[ {\dot{G}}_{K}^{T}C_{K}^{-1}{\dot{G}}_{K}\right] ^{T}\left[ K{\dot{G}}_{K}^{T}C_{K}^{-1}{\dot{G}}_{K}\right] ^{-1}\left[ {\dot{G}}_{K}^{T}C_{K}^{-1}{\dot{G}}_{K}\right] \right\} ^{-1}\nonumber \\ =&\left[ {\dot{G}}_{K}^{T}C_{K}^{-1}{\dot{G}}_{K}\right] ^{-1} \le \left[ {\dot{G}}_{K}^{T}Z^{T}\left( ZC_{K}Z^{T}\right) ^{-1}Z{\dot{G}}_{K}\right] ^{-1}, \end{aligned}
    $$
    which shows that $\Sigma_Q \le \Sigma_G$.

	To prove the optimality of the QIF estimator $\hat{\bm\beta}_Q$ in (d), we will show that the asymptotic covariance of $\hat{\bm\beta}_Q$ reaches the minimum among those of $\bm{\hat{\beta}}_E$ from $U^E(\bm{\beta})=0$.
	We consider that a matrix $A$ is minimized if $B-A$ is positive definite for any matrix $B$ with the same dimensions as matrix $A$.

    Similar to the discussion in Section \ref{section:2.2}, solving a class of estimating equations $ U^{E}(\bm{\beta})=0$ could be considered as minimizing a quadratic function, i.e.,
    $$ \begin{aligned}\bm{\hat{\beta}}_E=\arg \min _{\bm{\beta} } G^{T}_{K}(\bm{\beta} )W_K^{-1}G_{K}(\bm{\beta} ),\end{aligned}
    $$
    where $G_K( \bm{\beta} )$ is the extended score vector, $W_K$ is an $m(p_X+1)\times m(p_X+1)$ positive definite matrix, and $W_K \overset{p}{\rightarrow }\ W_0$ which is also an $m(p_X+1)\times m(p_X+1)$ positive definite matrix. Following the normality of the GMM estimator \citep{Hansen1982Eco}, we know that the asymptotic covariance matrix of $\hat{\bm{\beta}}_E$ is
    $$
    \begin{aligned} \Sigma_E(\hat{\bm{\beta}}) = \left( {\dot{G}}^T_K W_0 {\dot{G}}_K \right) ^{-1} {\dot{G}}^T_K W_0 \Omega W_0 {\dot{G}}_K ({\dot{G}}^T_K W_0 {\dot{G}}_K)^{-1}, \end{aligned}
    $$
    where $\Omega$ is the covariance matrix of $G_K$. We next show that $\Sigma_E(\hat{\bm{\beta}})$ has a lower bound $({\dot{G}}^T_K \Omega ^{-1} {\dot{G}}_K)^{-1},$ i.e.,
    $$
    \begin{aligned} \Sigma_E(\hat{\bm{\beta}})=\left( {\dot{G}}^T_K W_0 {\dot{G}}_K \right) ^{-1} {\dot{G}}^T_K W_0 \Omega W_0 {\dot{G}}_K \left( {\dot{G}}^T_K W_0 {\dot{G}}_K \right) ^{-1}\ge \left( {\dot{G}}^T_K \Omega ^{-1} {\dot{G}}_K \right) ^{-1}, \end{aligned}
    $$
    which is equivalent to show that
    $$
    \begin{aligned} S ={\dot{G}}^T_K \Omega ^{-1} {\dot{G}}_K- \left( {\dot{G}}^T_K W_0 {\dot{G}}_K \right) \left( {\dot{G}}^T_K W_0 \Omega W_0 {\dot{G}}_K \right) ^{-1} \left( {\dot{G}}^T_K W_0 {\dot{G}}_K \right) \ge 0, \end{aligned}
    $$
    that is, $S$ is a non-negative definite matrix. In fact, $S$ could be written as
    $$
    \begin{aligned}S ={\dot{G}}^T_K \Omega ^{-\frac{1}{2}} \{ I-\Omega ^{\frac{1}{2}}W_0{\dot{G}}_K [(\Omega ^{\frac{1}{2}}W_0{\dot{G}}_K)^{T} \Omega ^{\frac{1}{2}}W_0{\dot{G}}_K]^{-1} (\Omega ^{\frac{1}{2}}W_0{\dot{G}}_K)^T \} ({\dot{G}}^T_K \Omega ^{-\frac{1}{2}})^T .\end{aligned}
    $$
    Denote  $ L = { I-\Omega ^{\frac{1}{2}}W_0{\dot{G}}_K [(\Omega ^{\frac{1}{2}}W_0{\dot{G}}_K)^{T} \Omega ^{\frac{1}{2}}W_0{\dot{G}}_K]^{-1} (\Omega ^{\frac{1}{2}}W_0{\dot{G}}_K)^T }$.
    Since  ${L}^T L = L $ ,
    indicating that $L$ is an idempotent matrix, we have
    $$
    \begin{aligned}S ={\dot{G}}^T_K \Omega ^{-\frac{1}{2}} {L}^T L [{\dot{G}}^T_K \Omega ^{-\frac{1}{2}} ]^T= \left( L[{\dot{G}}^T_K \Omega ^{-\frac{1}{2}}]^T \right) ^T \left( L[{\dot{G}}^T_K \Omega ^{-\frac{1}{2}}]^T \right) \ge 0\end{aligned}
    $$
    and $S =\bm{0}$ if $W_0=\Omega ^{-1}$. In other words, when $W_0^{-1}$  is the covariance matrix of $G_K$, $\Sigma_E(\hat{\bm{\beta}})$ reaches the lower bound $({\dot{G}}^T_K \Omega ^{-1} {\dot{G}}_K)^{-1}$. We showed  
    that $\Sigma_Q(\hat{\bm{\beta}})=({\dot{G}}^T_K C_{K}^{-1} {\dot{G}}_K)^{-1}$ where $C_{K}=\frac{1}{K}\sum _{i=1}^{K}g_{i}(\bm\beta )g_{i}^{T}(\bm\beta)$. Since the sample covariance $C_{K}$ converges to $\Omega$ in probability, and ${\dot{G}}_K$ is nonrandom, so $({\dot{G}}^T_K C_{K}^{-1} {\dot{G}}_K)^{-1}$  converges in probability to $({\dot{G}}^T_K \Omega ^{-1} {\dot{G}}_K)^{-1}$. Therefore, the asymptotic covariance of $\hat{\bm\beta}_Q$, i.e., $\Sigma_Q(\hat{\bm{\beta}})$, reaches the minimum among $\Sigma_E(\hat{\bm{\beta}})$ of $\bm{\hat{\beta}}_E$ from $U^E(\bm{\beta})=0$.

    \bibliographystyle{abbrvnat}

    \bibliography{Refnew}

\end{document}